\documentclass[notitlepage,superscriptaddress,nofootinbib,tightenlines,twocolumn]{revtex4-1}
\usepackage{amsmath,verbatim,latexsym,amssymb,indentfirst,mathrsfs,mathtools,amsthm,bbm,bm,hyperref,url,cancel,subcaption}
\usepackage[font=small,labelfont=bf,format=plain,justification=raggedright,singlelinecheck=false]{caption}
\usepackage{verbatim,indentfirst}
\usepackage[title,titletoc]{appendix}

\usepackage{graphicx}
\usepackage{epstopdf}
\newcommand{\caphead}[1]{{\bf #1}}

\renewcommand{\thesection}{\Roman{section}}
\renewcommand{\thesubsection}{\Roman{section} \Alph{subsection}}
\renewcommand{\thesubsubsection}{\Roman{section} \Alph{subsection} \arabic{subsubsection}}
\makeatletter
\def\p@subsection{}
\makeatother
\makeatletter
\def\p@subsubsection{}
\makeatother

\makeatletter
\newcommand\footnoteref[1]{\protected@xdef\@thefnmark{\ref{#1}}\@footnotemark}
\makeatother

\usepackage{soul}

\usepackage{color}
\usepackage[dvipsnames]{xcolor}
\usepackage[normalem]{ulem}

\newcommand{\elec}{ {\rm elec} }

\newcommand{\Bath}{ \mathcal{B} }  
\newcommand{\Laser}{ {\rm laser} }  
  
\newcommand{\Energy}{ \mathcal{E} }
\newcommand{\EBath}{ \mathscr{E} }
\newcommand{\Decohere}{ \mathcal{D} }  
\newcommand{\Lind}{ \mathcal{L} }

\newcommand{\initial}{ {\rm i} }  
\newcommand{\final}{ {\rm f} }

\newcommand{\tick}{ {\rm tick} }

\newcommand{\mol}{ {\rm mol} }

\newcommand{\Fisher}{ \text{F} }

\newcommand{\TO}{ \mathcal{E} }

\newcommand{\LZ}{ {\rm LZ} }

\newcommand{\omegaP}{{(\omega)}}
\newcommand{\anc}{{\rm a}}
\newcommand{\Moment}{\mathcal{I}}

\newcommand{\Min}{ {\rm min} }   
\newcommand{\Max}{ {\rm max} }   
\newcommand{\hc}{ {\rm h.c.} }

\newcommand{\Sh}{{\rm Sh}}   
\newcommand{\Tr}{{\rm Tr}}   
\def\id{\mathbbm{1}}   

\newcommand{\Hil}{\mathcal{H}}  
\newcommand{\Sys}{\mathcal{S}}  
\newcommand{\Dim}{d}   

\newcommand{\LParen}{ \bm{(} }
\newcommand{\RParen}{ \bm{)} }

\newcommand*{\Set}[1]{\left\{  #1  \right\}}

\newcommand*{\bra}[1]{\langle #1\rvert}
\newcommand*{\ket}[1]{\lvert #1 \rangle}

\newcommand*{\ketbra}[2]{\lvert #1 \rangle\!\langle #2 \rvert}
\newcommand*{\expval}[1]{\left\langle  #1  \right\rangle}

\begin{document}
\title{Fundamental limitations on photoisomerization 
from thermodynamic resource theories}
\author{Nicole~Yunger~Halpern\footnote{E-mail: nicoleyh@g.harvard.edu}}
\affiliation{Institute for Quantum Information and Matter, California Institute of Technology, Pasadena, CA 91125, USA}
\affiliation{Kavli Institute for Theoretical Physics, University of California, Santa Barbara, CA 93106, USA}
\affiliation{ITAMP, Harvard-Smithsonian Center for Astrophysics, Cambridge, MA 02138, USA}
\affiliation{Department of Physics, Harvard University, Cambridge, MA 02138, USA}
\affiliation{Research Laboratory of Electronics, Massachusetts Institute of Technology, Cambridge, Massachusetts 02139, USA }
\author{David~T.~Limmer\footnote{E-mail: dlimmer@berkeley.edu}}
\affiliation{Department of Chemistry, University of California, 
Berkeley, CA 94720, USA}
\affiliation{Kavli Energy NanoSciences Institute, University of California, 
Berkeley, CA 94720, USA}
\affiliation{Lawrence Berkeley National Laboratory, University of California, Berkeley, CA 94720, USA}
\date{\today}

\begin{abstract}
Small, out-of-equilibrium, and quantum systems defy simple thermodynamic expressions. Such systems are exemplified by molecular switches, which exchange heat with a bath. These molecules can photoisomerize, or change conformation, or switch, upon absorbing light. The photoisomerization probability depends on kinetic details that couple the molecule's energetics to its dissipation. Therefore, a simple, general, thermodynamic-style bound on the photoisomerization probability seems out of reach. We derive such a bound using a resource theory. The resource-theory framework is a set of mathematical tools, developed in quantum information theory, used to generalize thermodynamics to small and quantum settings. From this toolkit has been derived a generalization of the second law, the thermomajorization preorder. We use thermomajorization to upper-bound the photoisomerization probability. Then, we compare the bound with an equilibrium prediction and with a Lindbladian model. We identify a realistic parameter regime in which the Lindbladian evolution saturates the thermomajorization bound. We also quantify the energy coherence in the electronic degree of freedom, and we argue that this coherence cannot promote photoisomerization. This work illustrates how quantum-information-theoretic thermodynamics can elucidate complex quantum processes in nature, experiments, and synthetics. 
\end{abstract}

{\let\newpage\relax\maketitle}

%
%
Thermodynamics quantifies ideal processes with simple expressions
and constrains processes that deviate from the ideal.
This simplicity vanishes in the face of small systems 
and intermediate time scales.
Such realistic settings yoke work, heat, and entropy production 
to kinematic details, exposing each as a fluctuating quantity.
Yet one can hope to bound kinematic results with 
general thermodynamic-style expressions.
Such results have enjoyed theoretical and experimental success. 
For example, fluctuation theorems and Jarzynski's equality~\cite{jarzynski1997nonequilibrium,crooks1999entropy,schuster2013nonequilibrium} 
constrain ensembles of irreversible transformations.
These results govern experiments,
including with single molecules~\cite{collin2005verification,trepagnier2004experimental}, 
and information engines~\cite{mandal2012work,toyabe2010experimental,koski2014experimental}. 
Additionally, thermodynamic uncertainty relations~\cite{barato2015thermodynamic,gingrich2016dissipation} 
have constrained the precision with which microscopic currents
can be generated.
These findings have relevance to molecular motors~\cite{barato2014efficiency,barato2016cost} 
and self-assembly~\cite{nguyen2016design,marsland2017limits}. 

We derive general, thermodynamic-style statements about 
small far-from-equilibrium quantum systems:
We derive bounds on molecular switches' photoisomerization,
or switching driven by light.
To do so, we leverage the quantum-information (QI) tool 
of thermodynamic resource theories. 

Small molecules photoisomerize, or switch configurations, in many
natural and synthetic systems~\cite{waldeck1991photoisomerization,Wang_94_Vibrationally,bandara2012photoisomerization,Hahn_00_Quantum,Hahn_02_Ultrafast,Browne_06_Making,Kahan_07_Following,levine2007isomerization,Johnson_17_Primary}.
Example photoisomers include retinal in rhodopsin, 
a pigment in the retina~\cite{strauss2005retinal}, 
and green fluorescent protein, 
a chromophore used throughout molecular imaging~\cite{tsien1998green}.
Three conveniences account for these molecular switches' prevalence.
First, ultrafast femtosecond lasers offer control over photoisomerization~\cite{Vogt_05_Optimal,Vogt_06_Analysis,brif2010control,prokhorenko2006coherent,arango2013communication}.
Second, photoisomers can be synthesized easily.
Third, one can easily encourage the expression of genes that code for isomers.
Applications are widespread and include
azobenzene-based solar-to-thermal fuels~\cite{Kucharski_14_Templated} 
and functional polymers~\cite{zhao2009smart,brif2010control}.

Despite their usefulness and prevalence,
photoisomers are not understood completely:
Experimental tools, such as time-resolved spectroscopy~\cite{zewail1996femtochemistry,mukamel2000multidimensional,jonas2003two,oliver2014correlating},
expose only a subset of the relevant degrees of freedom (DOFs).
Additionally, computational tools for simulating these processes 
remain under active development~\cite{micha2012density,kelly2013efficient,montoya2015extending,miller1970classical,ben1998nonadiabatic,topaler1996path,kapral1999mixed,mclachlan1964variational,tully1990molecular}.
We will help remedy the deficiency using QI theory.

\begin{figure}[t]
\centering
\includegraphics[width=8.2cm]{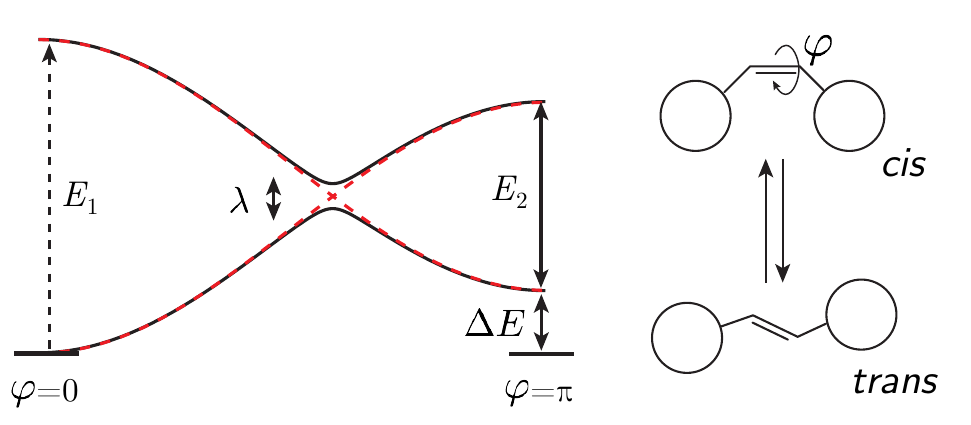}
\caption{\caphead{Photoisomer's energy landscape and conformations.}
Two representative potential-energy surfaces 
for the ground and excited electronic states of an isomer, 
as well as the \emph{cis} and \emph{trans} configurations 
associated with the ground-state minima. 
The black curves represent adiabatic states,
or instantaneous energy eigenstates.
The dashed red curves represent diabatic states,
which approximately equal adiabatic states at $\varphi = 0, \pi$.}
\label{Fig:1}
\end{figure}

Figure~\ref{Fig:1} depicts a photoisomer and its energy landscape.
Photoisomerization, or switching, begins with
the molecule in its thermal state.
The lowest electronic energy eigenstate approximates
the room-temperature thermal state well.
The heavy atoms' positions are parameterized with an angle $\varphi$.
$\varphi$ determines the electronic energy levels,
because the electronic energy spacing far exceeds
the energy spacings associated with
the molecule's vibrations and rotations~\cite{nitzan2006chemical}. 
The ground electronic level exhibits two wells,
centered near $\varphi = 0$ and $\varphi = \pi$.
An energy barrier separates the well.
$\varphi = 0, \pi$ define two conformations, or isomers, of the molecule:
the \emph{cis} and \emph{trans} configurations. 
Low-lying excited electronic states can lack energy barriers.
If excited by light, therefore,
the molecule has the opportunity to change configurations 
while relaxing, in contact with its environment, to the lower electronic level.
The probability of changing conformation during relaxation
is called the ``photoisomerization yield.''


The yield is difficult to predict for several reasons.
First, dynamical factors determine the yield
over intermediate time scales. 
These times 
exceed 
the time needed for the electronic DOF to relax to its ground state 
but are shorter than
the time over which the whole molecule thermalizes. 
This intermediacy precludes 
straightforward thermodynamic statements.
Second, the postexcitation evolution involves 
nonadiabaticity~\cite{tully2012perspective}, dissipation~\cite{tully2012perspective}, 
and rare bath fluctuations~\cite{schile2018studying}.

Hence studying the evolution computationally is difficult,
and few general guiding principles exist. 
We need a toolkit for deriving 
thermodynamic-style bounds on photoisomerization.
These bounds should incorporate
the coupling of quantum mechanical DOFs
with small scales and thermal fluctuations.
To construct such bounds, we 
turn to QI theory.

Resource theories are simple models
developed in QI theory~\cite{Coecke_16_Mathematical,Chitambar_18_Quantum}.
They are relevant when restrictions constrain
the processes that can occur, 
called ``free operations,''
and the systems accessible, called ``free systems.'' 
Consider, as an example,
a thermodynamic setting in which
systems exchange heat with a bath at a fixed temperature.
The first law of thermodynamics 
constrains processes to conserve energy,
and only thermal states
can be accessed easily.
The corresponding resource theory's free operations 
are called ``thermal operations.''
All nonfree systems, e.g., systems not in states
thermal with respect to the environment's temperature, are ``resources.''
Resources have value because 
they can fuel tasks such as work extraction.
Resource theories originated to quantify entanglement 
and to clarify which QI-processing tasks entanglement could facilitate~\cite{Horodecki_09_Quantum}.
Since then, resource theories have been developed 
for other valuable quantities, including
reference frames~\cite{vanEnk_05_Quantifying,Vaccaro_08_Tradeoff,Bartlett_06_Degradation,Bartlett_07_Reference,Gour_08_Resource,Marvian_13_Theory}, 
randomness used in cryptography~\cite{Coecke_16_Mathematical}, 
coherence~\cite{Baumgratz_14_Quantifying,Winter_16_Operational,Chitambar_16_Critical},   
``magic states'' used in quantum computation~\cite{Veitch_14_Resource}, and
thermodynamics~\cite{Lieb_91_Physics,Janzing_00_Thermodynamic,Horodecki_03_Local,Brandao_13_Resource,Horodecki_13_Fundamental,Gour_15_Resource,NYH_18_Beyond,Narasimhachar_15_Low,Lostaglio_15_Description}.

Using a resource theory, one studies 
which systems $R$ can transform into systems $S$
under free operations ($R \mapsto S$);
which cannot ($R \not\mapsto S$);
how much of a resource $W$, such as work, is required
to facilitate an otherwise impossible transformation
($R + W \mapsto S$ despite $R \not\mapsto S$);
how many copies of $S$ can be extracted from $m$ copies of $R$;
and what, generally, is possible and impossible.
Results govern arbitrarily small systems
and coherent quantum states.
In thermodynamic resource theories, 
averaging in a large-system limit
reproduces results consistent with 
expectations from statistical mechanics.
Hence resource theories offer the potential for formulating
sharp, general statements about complex, quantum systems.
We harness this potential for 
molecules undergoing photoisomerization.

This paper is organized as follows.
First, we review the resource theory that models heat exchanges.
We then model the molecule within the resource theory
(Sec.~\ref{sec_Model}).
We bound the photoisomerization yield
in two steps (Sec.~\ref{sec_Isomerization_Bound}):
First, we show that the electronic state's coherences
relative to the energy eigenbasis cannot promote isomerization,
if the environment is Markovian.
This result enables us to bound the isomerization yield
by focusing, second, on the density matrix's populations:
We apply thermomajorization, a resource-theory result
that generalizes the second law of thermodynamics.
The yield is tightly constrained, we find, 
if the light source barely excites the molecule,
as in some experiments. 
If the light source fully excites the molecule to 
one high-energy eigenstate, 
thermomajorization's upper bound on the yield
can be saturated transiently. 
%

We next quantify the coherences,
relative to the energy eigenbasis,
that the electronic state gains during photoisomerization
(Sec.~\ref{sec_Post_Isom_Coh}).
This coherence emerges after the molecule undergoes
a dissipative Landau-Zener evolution,
which we model within the resource theory.
The model indicates that the molecule acts as a quantum clock~\cite{Pauli_33_Handbuch,Pauli_58_Encyclopedia,Feynman_86_Quantum,Margolus_86_Quantum,Aberg_14_Catalytic,Malabarba_15_Clock,Frenzel_16_Quasi,Woods_16_Autonomous,Erker_17_Autonomous,Woods_18_Quantum}.
An application of this coherence appears in the appendices,
where work is shown to be extractable from the coherence.
Also, the work required to photoexcite the molecule to an arbitrary electronic state
in a single shot is calculated in the appendices.
The Discussion concludes with 
this program's significance and opportunities 
(Sec.~\ref{sec_Outlook}).

%
%
%
\section{Resource-theory model \\ for the photoisomer}
\label{sec_Model}

In this section, we review the resource theory 
that models heat exchanges.
Then, we model the chemical problem within the resource theory:
the molecule, bath, and light source in Sec.~\ref{sec_RT_Model_Isomer},
followed by photoisomerization in Sec.~\ref{sec_Model_Isomerization}.
To specify a system in the resource theory, 
one specifies a tuple $(\rho, H)$.
The $\rho$ denotes a quantum state,
and the $H$ denotes a Hamiltonian.
Both are defined on a Hilbert space $\Hil$ of dimensionality $\Dim$.

Each thermal operation consists of three steps:
(i) A thermal system governed by an arbitrary Hamiltonian $H_\Bath$
is drawn from the bath at inverse temperature $\beta$:
$( \rho_\Bath {=} \exp [-\beta H_\Bath ] / Z_\Bath,  \:
          H_\Bath )$,
wherein  $Z_\Bath := \Tr_\Bath [ \exp ( - \beta H_\Bath ) ]$ 
denotes the partition function.
(ii) The system and bath interact via 
an arbitrary energy-conserving unitary $U$.
(iii) A generalized environment $\Bath'$ is discarded.
$\Bath'$ is often the bath, $\Bath$,
but may be another subsystem.
Mathematically, a thermal operation $\mathcal{T}$ is represented as
\begin{align}
   \label{eq:Therm_Op}
   (\rho,    H)
   & \mapsto \mathcal{T}(\rho,    H) \equiv  
   \Big(  \Tr_{\Bath'}  \left\{  U
            \left[  \rho  \otimes  \rho_\Bath  \right]
            U^\dag  \right\} ,  
            H   
   \Big) .
\end{align}
The unitary satisfies a manifestation of the first law:
\begin{align}
   \label{eq_U_Restrict}
   [U,  H  +  H_\Bath]  =  0 .
\end{align}
The Hamiltonians are composed as 
$H + H_\Bath 
\equiv  (H  \otimes  \id)  +  (\id  \otimes  H_\Bath)$.

Only Eq.~\eqref{eq_U_Restrict} restricts $U$.
We do not assume that the system-bath interaction has 
any particular form or coupling strength;
we assume only that \eqref{eq_U_Restrict} is satisfied.
$U$ can transfer arbitrary amounts of energy
between the system and the bath in arbitrary times.
To test our resource-theory results, we compare them with Lindblad models in Sections~\ref{sec_Kinetic} and~\ref{sec_Dissip_LZ}. The Lindblad models involve more-specific couplings between the photoisomer and the bath. Such specific couplings are not required for most of our results.
To model photoisomerization, we focus on an $H$ that governs a photoisomer.

%
%
%
\subsection{Resource-theory model for \\
the molecule, bath, and light source}
\label{sec_RT_Model_Isomer}

\textbf{Resource-theory model for the molecule:}
The angular DOF $\varphi \in [0,  \pi]$
parameterizes the isomer's configuration
and governs the electronic Hamiltonian~\cite{Hahn_02_Ultrafast}.
We attribute to the molecule the Hamiltonian
\begin{align}
   H_\mol  
   & :=  \int_0^\pi  d \varphi  \;
   H_\mol (\varphi) 
   \label{eq:H_Mol} \\
   & \equiv   \int_0^\pi  d \varphi  \;  \left[
   H_\elec ( \varphi)  \otimes
   \ketbra{\varphi}{\varphi}
   +  \id_\elec  \otimes  
   \frac{ \ell_\varphi^2 }{2 \Moment}  \right]  \, \nonumber .
\end{align}
In each term in Eq.~\eqref{eq:H_Mol},
the first factor acts on
an electronic Hilbert space $\Hil_\elec$,
and the second acts on
a configurational Hilbert space $\Hil_\varphi$.
$\id_\elec$ and $\id_\varphi$ denote the identity operators
on $\Hil_\elec$ and on $\Hil_\varphi$.
$\ell_\varphi$ denotes the angular-momentum operator
associated with the quasiclassical mode $\varphi$,
which has a moment of inertia $\Moment$.
The angular DOF is well-approximated as quasiclassical
due to the chemical groups' sizes and masses~\cite{Wang_94_Vibrationally}:
The groups localize at angular coordinates
far from the $\varphi$ values at which
$H_\elec (\varphi)$ is degenerate,
satisfying the Born-Oppenheim approximation.

Our $H_\mol$ has the form of Hamiltonians in~\cite{Horodecki_13_Fundamental,Malabarba_15_Clock}.
There, a switch changes the system-of-interest Hamiltonian. 
$\varphi$ acts as the switch here,
and the electronic DOF acts as the system.
A related model appeared very recently in~\cite{Pusuluk_18_Proton}.

The electronic Hamiltonian has the form in~\cite{Hahn_02_Ultrafast}:
\begin{align}
   \label{eq:H_Elec_Hahn}
   H_\elec (\varphi) 
   =&  \left[  (2 E_1 -\Delta E - E_2) \sin^2 (\varphi/2)  \right]
   \ketbra{ \psi_0 }{ \psi_0 }
   \\ \nonumber & 
   +  \left[  E_1  -  (E_1-\Delta E) \sin^2 (\varphi/2)  \right]
   \ketbra{\psi_1}{\psi_1}
   \\ \nonumber &
   +  \frac{\lambda}{2}  ( \ketbra{ \psi_0}{\psi_1}  +  \ketbra{ \psi_1}{\psi_0} )  \, .
\end{align}
Figure~\ref{Fig:1} illustrates the energy landscape.
The diabatic basis $\Set{ \ket{ \psi_0 },  \ket{\psi_1} }$
approximately diagonalizes the Hamiltonian
at $\varphi = 0, \pi$.
The constants $E_1, E_2, \Delta E  > 0$
far exceed the interstate coupling 
$\lambda  >  0$.
The energy required to excite  
$\ket{\psi_0}   \ket{ \varphi{=}0 }$  to 
$\ket{\psi_1}   \ket{ \varphi{=}0 }$ 
is denoted by $E_1$.
The energy of excitation from $\ket{\psi_0}   \ket{ \varphi{=}\pi }$  to 
$\ket{\psi_1}   \ket{ \varphi{=}\pi }$  is $E_2$.
The energy stored during a transition from 
$\ket{\psi_0}   \ket{ \varphi{=}0 }$ to 
$\ket{\psi_0}   \ket{ \varphi{=}\pi }$ is 
$\Delta E$. 

We notate the eigenenergies by $\Energy_\pm (\varphi)$,
such that $\Energy_+ (\varphi)  \geq  \Energy_- (\varphi)$,
and the adiabatic basis by 
$\Set{  \ket{ \Energy_\pm (\varphi) }  }$:
\begin{align}
   \label{eq_Diag_H_Elec}
   H_\elec (\varphi)
   =  \sum_{ \mu = \pm }
   \Energy_\mu (\varphi)
   \ketbra{ \Energy_\mu (\varphi) }{ \Energy_\mu (\varphi) } .
\end{align}

Though simple, this single-mode model reproduces
linear and time-dependent spectroscopy of photoswitches like rhodopsin.
The model also accounts for high reaction efficiencies~\cite{Hahn_00_Quantum}. 
Furthermore, the model captures the photoisomerization yield's
dependence on environmental factors~\cite{Kim_03_Analysis,Hahn_02_Ultrafast,Burghardt_06_Excited}.
Consequently, models of the form of Eq.~\ref{eq:H_Elec_Hahn} 
are routinely studied to explore environmental effects~\cite{stock2005classical,brunk2015mixed}
and to benchmark novel numerical techniques~\cite{worth2008using,arango2013communication,Videla_18_Floquet,Sala_18_Ultrafast}.

%
\textbf{Resource-theory model for the bath:}
A Hamiltonian $H_\Bath  =  \sum_k  \EBath_k  
\ketbra{ \EBath_k }{ \EBath_k }$ 
governs the bath, which occupies a Gibbs state
$\rho_\Bath
= \sum_k \exp(- \beta \EBath_k) / 
Z_\Bath \ketbra{ \EBath_k }{ \EBath_k }$.
We assume that $H_\Bath$ has the properties
in~\cite[Suppl.~Note~1, p.~5]{Horodecki_13_Fundamental},
invoking the justifications therein.
For example, degeneracies are assumed to grow exponentially with energy.

%
\textbf{Resource-theory model for the light source:}
A laser or sunlight performs work on the molecule~\cite{Aberg_14_Catalytic,Lostaglio_15_Description,Lostaglio_15_Quantum,Lostaglio_18_Elementary,Bedkihal_16_Comment}.
We model the light as a multimode bosonic field in a state $\rho_\Laser$.
Examples include coherent states, 
but $\rho_\Laser$ can have an arbitrary form.
$\rho_\Laser$ dictates to which states the molecule can be photoexcited,
and our results hold for arbitrary photoexcited states.
We consider a range of photoexcited states
(Sections~\ref{sec_Thermomaj_Bound} and~\ref{sec_Kinetic}) 
and so a range of $\rho_\Laser$'s.

We denote the oscillator Hamiltonian by $H_\Laser$ and expand it in 
particle-number states $\ket{ n_{ \omega } }$
that satisfy the eigenvalue equation
$N_{ \omega }  \ket{ n_{ \omega } } 
=  n_\omega  \ket{ n_{ \omega } }$,
wherein $n_\omega = 0, 1, 2, \ldots$
Our arguments do not rely on the precise form of 
the density of states.

Each fixed-$\omega$ term resembles the ladder Hamiltonians
with which batteries have been modeled in
thermodynamic resource theories~\cite{Skrzypczyk_13_Extracting,Skrzypczyk_14_Work,NYH_16_Beyond,NYH_18_Beyond,Aberg_14_Catalytic}.
Batteries are often modeled with spectra unbounded from below,
because ground states can complicate accountings of coherence~\cite{Masanes_14_Derivation,Aberg_14_Catalytic}.
$H_\Laser$ has a ground state, 
modeling a real physical Hamiltonian.
However, the ground state will lack much population.
Hence comparing the light source with resource-theory batteries is justified.



%
%
%
\subsection{Resource-theory model for photoisomerization}
\label{sec_Model_Isomerization}

The molecule begins in thermal equilibrium with the bath,\footnote{
In certain cases, the molecule may begin out of equilibrium~\cite{Wang_94_Vibrationally,Kahan_07_Following,Johnson_17_Primary}. A source of work and/or coherence keeps the molecule from thermalizing. This resource can straightforwardly be incorporated into our model. Such a resource would affect the postexcitation states accessible to the molecule (the right-hand side of Eq.~\eqref{eq_Excite_Th_Op}). Our main result, the thermomajorization bound on the photoisomerization yield, governs arbitrary postexcitation states. Hence a nonequilibrium initial state does not affect our result's form.
}
in the state $\rho = \exp( - \beta H_\mol ) / Z_\mol$. 
This state follows from long-time thermalization.
We will focus on the molecule's later rotation,
which happens over a short time 
and can break the weak-coupling assumption that leads to thermal states.
We assume that Eq.~\ref{eq:H_Elec_Hahn} 
is parameterized such that
the \emph{cis} isomer is strongly energetically preferred~\cite{Hahn_02_Ultrafast}:
$\rho  \approx  \ketbra{ \psi_0 }{ \psi_0 }  \otimes
\ketbra{ \varphi{=}0 }{ \varphi{=}0 }$.
Our results extend straightforwardly to isomers
whose \emph{trans} states are preferred, such as azobenzene.

We model photoisomerization with three thermal operations,
represented by arrows $\mapsto$.
First, the laser excites the molecule,
\begin{align}
   \label{eq_Excite_Th_Op}
   &  e^{ - \beta H_\mol } / Z_\mol 
   \otimes  \rho_\Laser
   \mapsto
   \rho_\initial   \otimes  
   \ketbra{ \varphi{=}0 }{ \varphi{=}0 }  \, ,
\end{align}
at the angular coordinate $\varphi = 0$.
The laser forms part of the generalized environment $\Bath'$
traced out in Eq.~\eqref{eq:Therm_Op}.
$\rho_\initial$ denotes the new electronic state,
which depends on $\rho_\Laser$ and on 
the thermal operation's form (on how the molecule couples to the bath and the laser).
Our calculations are evaluated on a range of $\rho_\initial$'s,
from fully to barely excited (Sections~\ref{sec_Thermomaj_Bound} and~\ref{sec_Kinetic}).
Second, the molecule rotates: 
\begin{align}
   \label{eq_TO2}
   \rho_\initial   \otimes  
   \ketbra{ \varphi{=}0 }{ \varphi{=}0 }
   \mapsto  \rho_\final  .
\end{align}
$\rho_\final$ denotes the post-photoisomerization state.
Most of its weight lies on the \emph{trans} states,
for which $\varphi = \pi$.
Maximizing the isomerization yield
amounts to maximizing the final state's weight 
on the lower \emph{trans} level,
\begin{align}
   \rho_-(\pi)  
   :=  \bra{ \Energy_- (\pi),  \varphi {=} \pi }  
   \rho_\final  
   \ket{ \Energy_- (\pi),  \varphi {=} \pi } .
\end{align}
Third, the molecule thermalizes to 
$\exp(- \beta H_\mol ) / Z_\mol$.

Appendix~\ref{sec:LZ_RT} details a more sophisticated model for step 2:
The thermal operation~\eqref{eq_TO2} is decomposed as
a sequence of thermal operations.
The angular DOF $\varphi$ serves as a quantum clock~\cite{Pauli_33_Handbuch,Pauli_58_Encyclopedia,Feynman_86_Quantum,Margolus_86_Quantum,Aberg_14_Catalytic,Malabarba_15_Clock,Frenzel_16_Quasi,Woods_16_Autonomous,Erker_17_Autonomous,Woods_18_Quantum}
as the molecule rotates at some speed $v$.
This sequence models a dissipative Landau-Zener transition.
The sophisticated model (of a sequence of thermal operations) 
is consistent with the simpler model (of one thermal operation):
Every composition of thermal operations is a thermal operation
(App.~\ref{sec_Compose_Th_Ops}).
Every thermal operation---including every composite---obeys thermomajorization.
To test our resource-theory results, we compare them with specific minimal models in Sections II D and III B. The models involve specific Lindbladian couplings between the photoisomer and the bath. However, such specific couplings are not required for most of our results.

%
%
%
\section{Limitations on \\photoisomerization yield}
\label{sec_Isomerization_Bound}

The rotational thermal operation~\eqref{eq_TO2}  
leaves the photoisomer in a state $\rho_\final$. 
We upper-bound the isomerization yield $\rho_{-}(\pi)$  
in two steps.
First, we show that the electronic state's coherences
relative to the energy eigenbasis
cannot affect the yield (Sec.~\ref{sec_Coher_Irrelevant}).
This result relies on the independence
of coherences' and populations' evolutions
under thermal operations~\cite{Lostaglio_15_Quantum,Marvian_14_Modes}.
The coherences' irrelevance enable us to bound the yield
by focusing on density matrices' populations.
The populations' evolutions are constrained by thermomajorization,
a preorder that generalizes the second law of thermodynamics~\cite{Ruch_75,Ruch_75_Mixing,Ruch_78_Information,Ruch_76_Principle,Ruch_78_Mixing,Janzing_00_Thermodynamic,Horodecki_13_Fundamental,Egloff_15_Measure,MarshallOA10}.
We review thermomajorization (Sec.~\ref{sec_Thermomaj_Backgrnd}),
then apply it to bound the photoisomerization yield
(Sec.~\ref{sec_Thermomaj_Bound}).
To evaluate the bound's tightness, 
we compare with Lindblad evolution (Sec.~\ref{sec_Kinetic}).
The bound is tight, we find, when kinetic parameters favor photoisomerization.

\subsection{Electronic energy coherences \\ cannot influence 
the photoisomerization yield}
\label{sec_Coher_Irrelevant}

In conventional thermodynamics,
a system's free energy declines monotonically
under spontaneous processes.
Resource-theory monotones behave similarly.
A monotone is a function $f$,
evaluated on a system $(\rho, H)$,
that decreases monotonically under free operations $\mathcal{T}$~\cite{MarshallOA10,Chitambar_18_Quantum,Gour_15_Resource}:
$f(\rho, H)  \geq  f \LParen \mathcal{T} (\rho, H) \RParen$.
Monotones quantify resourcefulness, 
which is eroded monotonically by free operations, i.e., thermalization.
Different monotones quantify 
the system's ability to fuel different tasks,
such as work extraction and timekeeping.

Coherence can be grouped into modes,
each associated with one energy gap~\cite{Lostaglio_15_Quantum,Marvian_14_Modes}.
Let $H = \sum_j  E_j  \ketbra{j}{j}$ denote a Hamiltonian
that governs a state $\rho$.
The $\omega$ mode of $H$
consists of the pairs $(j, k)$ whose gaps
$E_j - E_k  =  \omega$.
If $\rho_{jk}  :=  \bra{ j }  \rho  \ket{k}$,
then $\rho_{jk}$ encodes coherence when $j \neq k$.
%
A state's mode-$\omega$ coherence has been quantified with
the one-norm~\cite{Lostaglio_15_Quantum},
defined as
$|| A ||_1  :=  \Tr  \left(  \sqrt{ A A^\dag  }  \right)$
for a matrix $A$.
Suppose that some thermal operation $\mathcal{T}$ maps
$(\rho, H)$ to $(\sigma, H)$.
The modes' one-norms decay monotonically and independently~\cite{Marvian_14_Modes,Lostaglio_15_Quantum}:
\begin{align}
   \sum_{j, k  : E_j - E_k  =  \omega}
   || \rho_{jk} ||_1
   \geq  \sum_{j, k  : E_j - E_k  =  \omega}
   ||  \sigma_{jk}  ||_1 
   \quad  \forall \omega.
\end{align}
The modes' independence follows from thermal operations'
commuting with time translations.
We review the proof in App.~\ref{sec_Review_Coher_Indep}.
This result relies on sequential thermal operations' modeling
interactions with a Markovian bath.
Markovianity is a common assumption
that models rhodopsin's photoisomerization~\cite{Thoss_06_Quantum}.

\begin{figure}[t]
\includegraphics[width=8.5cm]{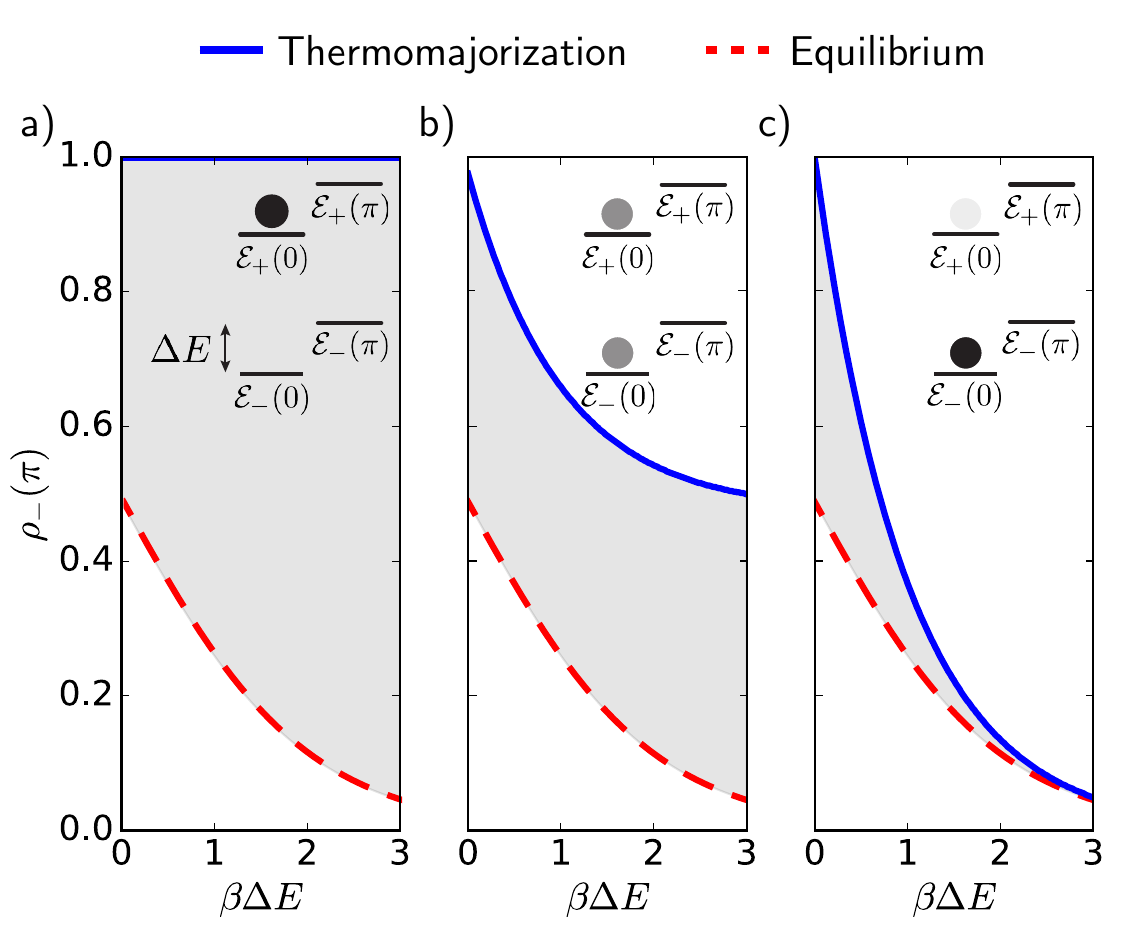}
\caption{\caphead{
Thermomajorization bound
on the photoisomerization yield $\rho_{-}(\pi)$
and comparisons with equilibrium statistical mechanics.}
The red dashed curve shows the predicted equilibrium yield, 
and the blue solid curve shows the resource-theory bound. 
Possible optimal yields (gray region) are shown for
an initially excited state a), 
an initial superposition b),
and an unexcited state c). 
The insets illustrate the molecule's energy levels.
The shaded dots show the initial state's probability weights.
}
\label{fig:Thermomaj_Bounds}
\end{figure}

The molecule's $H_\elec(0)$ has 
a coherence mode $\omega_1  =  \Energy_+ (0)  -  \Energy_- (0)= E_1$
and a population mode $\omega_0 = 0$.
The initial state $\approx \ket{ \Energy_- (0) }$
lacks coherence, 
so the laser provides all the coherence in 
the photoexcited state $\rho_\initial$.
Suppose, for example, that the laser creates an even superposition, 
$\frac{1}{ \sqrt{2} } \:
\LParen \ket{ \Energy_- (0) }  +  \ket{ \Energy_+ (0) } \RParen
\ket{ \varphi {=} 0 }$,
as in Fig.~\ref{fig:Thermomaj_Bounds}b).
Photoexcitation gives the state a nonzero amount
\begin{align}
   \label{eq:Coh_From_Laser}
   \left\lvert  \left\lvert   \sum_{(j, k) = (+, -), (-, +)}
   ( \rho_\initial )_{jk}
   \right\rvert   \right\rvert_1
   =  1
\end{align}
of coherence.
Since modes transform independently under thermal operations,
the $\omega_1$ electronic coherence
cannot influence the $\omega_0$ populations.
If the target \emph{trans} state is an energy eigenstate, therefore,
injecting coherence into the electronic state via photoexcitation  
cannot improve the isomerization yield.

%
%
%
\subsection{Background: Thermomajorization}
\label{sec_Thermomaj_Backgrnd}

The coherences' irrelevance frees us 
to focus on density matrices' populations
when bounding the photoisomerization yield.
The thermomajorization preorder governs 
the populations' evolution under thermal operations~\cite{Ruch_75,Ruch_75_Mixing,Ruch_78_Information,Ruch_76_Principle,Ruch_78_Mixing,Janzing_00_Thermodynamic,Horodecki_13_Fundamental,Egloff_15_Measure,MarshallOA10}.
Thermomajorization implies necessary and sufficient conditions
for the existence of a thermal operation 
that maps one energy-diagonal state into another.\footnote{
If the states are not energy-diagonal,
a generalization of thermomajorization 
encapsulates the necessary and sufficient conditions~\cite{Gour_18_Quantum}.
Bounding photoisomerization does not require the generalization:
The coherences cannot affect the photoisomerization yield,
which is one of the populations 
(Sec.~\ref{sec_Coher_Irrelevant}).}

Thermomajorization can be characterized as follows.
Let $H = \sum_{j = 1}^\Dim E_j \ketbra{j}{j}$ 
denote a Hamiltonian that governs 
a state $\rho$ of energy diagonal
$\Decohere(\rho)  
:=  \sum_j  \ketbra{j}{j}  \rho  \ketbra{j}{j}
=  \sum_j  r_j  \ketbra{j}{j}$.
Consider rescaling the probabilities with Boltzmann factors,
$r_j e^{ \beta E_j }$, 
and ordering the products from greatest to least:
$r_{j'} e^{ \beta  E_{j'} }  
\geq  r_{k'}  e^{ \beta E_{k'} }$
for all $j'  >  k'$.
Consider the points
$\left(  \sum_{j' = 1}^{ \alpha }  e^{ - \beta E_{j'} },
\sum_{j' = 1}^\alpha  p_{j'}  \right)$,
for all $\alpha = 1, 2, \ldots, \Dim$.
Connecting them with straight lines
defines a piecewise-linear curve.
This Gibbs-rescaled Lorenz curve is denoted by 
$L_{( \rho, H )}(x)$, 
wherein the $x$-coordinates run from 0 to
the partition function,
$Z  :=  \sum_{ j = 1}^\Dim  \exp( - \beta E_j )$.
%
Let $(\sigma, H)$ denote another system,
represented by a Lorenz curve $L_{ (\sigma, H) } (x)$. 
If the $(\rho, H)$ curve lies above or on the $(\sigma, H)$ curve
at all $x$-values, $(\rho, H)$
\emph{thermomajorizes} $(\sigma, H)$.
If and only if $(\rho, H)$ thermomajorizes $(\sigma, H)$
does some thermal operation map
the first state's energy diagonal to the second state's:
\begin{align}
   \label{eq_Lorenz_Law}
   & L_{(\rho, H)}(x)  \geq  L_{ (\sigma, H)}(x)
   \quad  \forall x  \in  [0, Z] 
   \quad \Leftrightarrow \\ \nonumber &
   \exists \mathcal{T} : 
   \mathcal{T} \LParen \Decohere (\rho), H \RParen  
   =  \LParen  \Decohere (\sigma), H  \RParen .
\end{align}
Relation~\eqref{eq_Lorenz_Law} generalizes 
the second law of thermodynamics
to arbitrarily small systems and to single-shot transformations.
The curve $L_{(\rho, H)}$ illustrates 
the thermodynamic value of $(\rho, H)$ 
by codifying the system's informational and energetic resourcefulness.


\subsection{Thermomajorization bound \\
on the photoisomerization yield}
\label{sec_Thermomaj_Bound}

To bound the optimal photoisomerization yield $\rho_{-}(\pi)$,
we construct the Gibbs-rescaled Lorenz curves for 
(i) the postexcitation state 
$\rho_\initial  \otimes  \ketbra{ \varphi {=} 0 }{ \varphi {=} 0 }$ and
(ii) the postrotation state $\rho_\final$.
We then solve for the greatest $\rho_{-}(\pi)$ that enables
the photoexcited state to thermomajorize $\rho_\final$.
The calculations were numerical; the code used is available at
\href{https://github.com/dlimmer/QIT-PhotoIsomer}{dlimmer/QT-PhotoIsomer}.

The study has the following focus.
We concentrate on the angles $\varphi=0,\pi$ 
that define the \emph{cis} and \emph{trans} configurations. 
Extensions of this four-level model are straightforward
and follow from our results, as discussed below.
We assess how the bound depends on 
the \emph{cis}-\emph{trans} energy gap 
$\Delta E := \Energy_- (\pi)-\Energy_- (0)$, 
expressed in units of $1/\beta$.
We focus on the physically relevant regime in which 
the ground-state-to-metastable-state gap
is far smaller than the \emph{cis} gap:
$\Delta E \ll \Energy_+ (0) - \Energy_- (0)
=: E_1$.
 
Figure~\ref{fig:Thermomaj_Bounds} shows results
for three photoexcited states $\rho_\initial$.
These states interpolate between two extremes,
the fully excited $\ket{ \psi_1 } \ket{ \varphi {=} 0 }$
and the ground state $\ket{ \psi_0 }  \ket{ \varphi {=} 0 }$.
If the laser fails to excite the molecule,
thermal excitations drive any isomerization.
In all three cases, we can identify kinetic setups
in which the photoisomerization yield saturates the resource-theory bound
(App.~\ref{sec_KineticMethods}).

We compare the resource-theory bound 
with the Boltzmann-factor yield
predicted by equilibrium statistical mechanics,
$\rho_{-}(\pi)  = \exp[ - \beta  \Energy_-(\pi)] / Z_\mol$. 
For all $\rho_\initial$, the equilibrium yield lies below 
the thermomajorization bound, as required. 
Moreover, the equilibrium yield lower-bounds the optimal yield:
Any additional kinetic preference for converting \emph{cis} to \emph{trans} increases the yield.
With thermomajorization upper-bounding possible outcomes, 
and equilibrium statistical mechanics lower-bounding them, 
we obtain a range of possible yields 
as a function of $\rho_\initial$ and $\Delta E$.


%

Suppose that the laser fully excites the molecule, to 
$\rho_\initial = \ketbra{\Energy_+(0)}{\Energy_+(0)}$.
Thermomajorization caps the yield trivially at one,
as shown in Fig.~\ref{fig:Thermomaj_Bounds}a).
Hence energy conservation and the fixed-temperature bath
do not limit the isomerization yield.
The unboundedness persists across the physically reasonable gaps 
$\Delta E \ll E_1$. 

If the laser half-excites the molecule, such that 
$\Decohere ( \rho_\initial )
= \frac{1}{2} \ketbra{\Energy_+(0)}{\Energy_+(0)}
+  \frac{1}{2} \ketbra{\Energy_-(0)}{\Energy_-(0)}$,
the yield obeys the bounds in
Fig.~\ref{fig:Thermomaj_Bounds}b).
When $\Delta E = 0$,
the thermomajorization bound $< 1$.
As $\Delta E$ grows,
the bound approaches $1/2$.
The bound remains $1/2$ for greater values of
$\Delta E$ than the plot shows.

Now, suppose that the laser fails to excite the molecule significantly:
$\rho_\initial = (1-\epsilon)\ketbra{\Energy_-(0)}{\Energy_-(0)} + \epsilon \ketbra{\Energy_+(0)}{\Energy_+(0)}$, wherein $\epsilon \ll 1$.
Thermal excitations drive the isomerization,
whose bounds are shown in Fig.~\ref{fig:Thermomaj_Bounds}c).
At large $\Delta E$,
the resource-theory bound asymptotes to $\epsilon$.
The bound approaches 1
as $\Delta E$ shrinks to 0.

Resource-theory insights explain several trends.
If the laser fully excites the molecule, we saw,
thermomajorization implies only that
$\rho_{-}(\pi)  \leq  1$.
The reason follows from how, as reflected in $L_{(\rho, H)}$,
thermodynamic resourcefulness decomposes 
into information and energy. 
The initial state is an energy eigenstate,
$\rho_\initial  =  \ketbra{ \Energy_+ (0) }{ \Energy_+ (0) }$, 
so an energy measurement's outcome is perfectly predictable. 
$\rho_\initial$ therefore encodes maximal information.
$\rho_\initial$ also has more energetic value than 
the lower \emph{trans} state, as
$E_1 > \Delta E$. 
Hence $\rho_\initial$ has far more resourcefulness than 
$\ket{ \Energy_-(\pi) }$.
The fundamental thermodynamic limitations 
of energy conservation and temperature
do not constrain the ability of $\rho_\initial$ 
to transform into $\ket{ \Energy_- (\pi) }$.
Only kinetic practicalities, such as relaxation rates, do. 

As probability weight shifts downward in $\rho_\initial$, 
$\rho_\initial$ loses energetic value.
$\rho_\initial$ loses also informational value:
The diagonal, $\Decohere (\rho_\initial)$, grows more mixed.
Hence the solid blue curve in Fig.~\ref{fig:Thermomaj_Bounds}b)
lies below the solid blue curve in Fig.~\ref{fig:Thermomaj_Bounds}a).
But the mixed $\Decohere (\rho_\initial)$ 
retains significant energetic value, 
since $E_1 \gg \Delta E$. 
If the laser fails to excite the molecule,
$\rho_\initial$ regains informational value,
being the energy eigenstate
$\ketbra{ \Energy_- (0) }{ \Energy_- (0) }$.
The dearth of energy outweighs 
this informational value, however. 

Our four-level model can be extended:
The isomer can begin or end in a probabilistic combination
of configurations $\varphi$.
The Gibbs-rescaled Lorenz curves will be calculated
by the same procedure.
The more the state's weight is distributed across configurations,
the less predictable an energy measurement,
and so the lesser the state's thermodynamic value.
Hence a distribution over initial configurations will lower the bound,
and a distribution over final configurations will tighten the bound.
Realistic mixtures of final states will be dominated by the
$\ket{\varphi{=}0}$ and $\ket{\varphi{=}\pi}$
on which we focus.

\subsection{Kinetic factors that saturate bound}
\label{sec_Kinetic}

Can the upper bounds be saturated? 
We gain insight from a minimal kinetic model 
of the post-photoexcitation relaxation.
We model the molecule's evolution with the Lindblad master equation
   $\dot{\rho}(t) = - \frac{i}{\hbar} [H,  \rho(t)] 
   + \Lind \LParen \rho (t)  \RParen $.
The first term represents the system's coherent dynamics. 
We again focus on two angular states, 
so that the minimal Hamiltonian is
$H \approx   \sum_{\varphi=0,\pi} 
  H_\elec (\varphi)$
[Eq.~\eqref{eq_Diag_H_Elec}].
We set $\Energy_- (0)=0$ and, as before, 
$\Energy_+ (0)=E_1$ and 
$\Energy_- (\pi)=\Delta E$. 
For simplicity, we take $\Energy_+ (\pi) = E_1 + \Delta E$, 
and we work in the physically relevant regime 
$E_1 \gg \Delta E>0$.
The levels' populations are denoted by
$p_\mu (\varphi)$.
The Lindblad equation's second term,
\begin{equation}
\label{Eq:Diss}
   \Lind(\rho) 
   = \sum_i \Gamma_i \left 
   (B_i \rho B^\dagger_i 
   - \frac{1}{2} \{B^\dagger_i B_i , \rho(t) \}\right ) ,
\end{equation}
reflects the influence of the bath, 
which decoheres the state and dissipates energy. 
We choose Lindblad operators  
$B_{\Energy_\mu(\varphi),\Energy_{\mu '}(\varphi')}
= \ketbra{ \Energy_\mu (\varphi) }{ \Energy_{\mu'} (\varphi')}$,  
for each pair of energy eigenstates.
Each $B_i$ dissipates at a rate  $\Gamma_i$
that satisfies local detailed balance,
so the system relaxes toward a thermal state. 
The parameters strongly kinetically prefer 
relaxation into the \emph{trans} state (App.~\ref{sec_KineticMethods}). 

\begin{figure}[b]
\includegraphics[width=8.5cm]{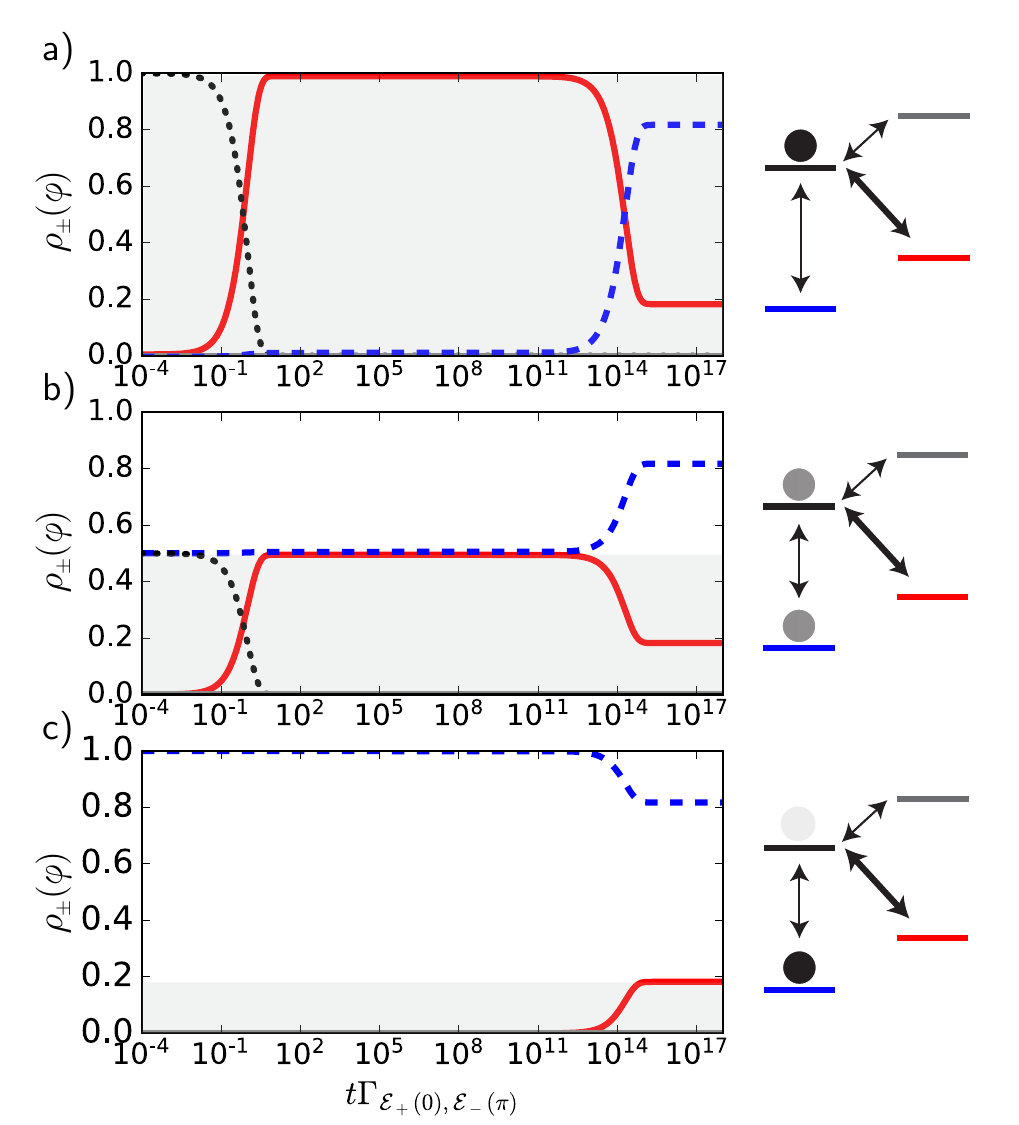}
\caption{\caphead{Comparison of thermomajorization bound 
with time-dependent Lindblad dynamics}. 
Calculations are performed on the four-level system 
shown on the right. 
The filled circles illustrate the initial probability weights,
and the arrows signify the possible transitions.
Each energy level's population evolves
as the correspondingly colored curves in the plots:
$\ket{\Energy_+(0)}$ (dotted black), 
$\ket{\Energy_-(\pi)}$ (solid red), 
$\ket{\Energy_-(0)}$ (dashed blue), and 
$\ket{\Energy_+(\pi)}$ (solid gray). 
The grayed area denotes the region accessible to 
$\ket{\Energy_-(\pi)}$ according to thermomajorization.
Population dynamics are shown following
a) full excitation to $\ket{\Energy_+(0)}$,
b) half-excitation to a state of energy diagonal
$\frac{1}{2} \ketbra{\Energy_+(0)}{\Energy_+(0)}
+  \ketbra{\Energy_-(0)}{\Energy_-(0)}$, and
c) failure to excite the state significantly: $\rho_\initial 
= (1-\epsilon)\ketbra{\Energy_-(0)}{\Energy_-(0)} 
+ \epsilon \ketbra{\Energy_+(0)}{\Energy_+(0)}$, 
wherein $\epsilon \ll 1$.
The parameters used are $\beta \Delta E=1.5$, 
$\beta E_1=30$, 
$\beta \hbar \Gamma_{\Energy_+(0),\Energy_-(\pi)}  =  1$, and 
$\beta  \hbar  \Gamma_{\Energy_+(0),\Energy_-(0)}
=  \beta  \hbar  \Gamma_{ \Energy_+(\pi),  \Energy_+ (0) }=0.01$.}
\label{fig:Kinetic_yields}
\end{figure}

Figure~\ref{fig:Kinetic_yields} shows the time-dependent densities 
following relaxation from 
the initial conditions in Sec.~\ref{sec_Thermomaj_Bound}.
Figure~\ref{fig:Kinetic_yields}a) follows from
the fully excited initial state
$\rho_\initial  =  \ketbra{ \Energy_+(0) }{ \Energy_+(0) }$,
and Fig.~\ref{fig:Kinetic_yields}b) follows from
the partially excited
$\rho_\initial 
= \frac{1}{2} \ketbra{\Energy_+(0)}{\Energy_+(0)}
+ \frac{1}{2}  \ketbra{\Energy_-(0)}{\Energy_-(0)}$.
In these cases, the thermomajorization bound 
is saturated at intermediate times, 
before thermalization reduces the yield to its equilibrium value. 
In contrast, Fig.~\ref{fig:Kinetic_yields}c) follows from
a barely excited initial state,
$\rho_\initial = (1-\epsilon)\ketbra{\Energy_-(0)}{\Energy_-(0)} 
+ \epsilon \ketbra{\Energy_+(0)}{\Energy_+(0)}$, wherein $\epsilon \ll 1$.
The yield maximizes, reaching the resource-theory bound,
at very long times set by the large barrier in the ground electronic state,
$\Energy_-(\varphi)$ (Fig.~\ref{Fig:1}).

\section{Limitations on coherence following photoisomerization}
\label{sec_Post_Isom_Coh}

Photoexcitation may inject coherence into the molecule's state. 
By ``coherence,'' we mean, 
coherence relative to the energy eigenbasis
in the electronic state. 
Such coherence might be expected, \emph{a priori},
to contribute to the molecule's rotation. 
Using the resource-theory tool of monotones, 
we bound the amount of coherence in $\rho_\final$. 
Our argument involves sequential thermal operations,
which model, in the resource theory, interactions with a Markovian bath.
Markovianity is a common assumption
that models rhodopsin's photoisomerization~\cite{Thoss_06_Quantum}.
This coherence has an application detailed in App.~\ref{sec_Work}:
Work can be extracted from the coherence, if photoisomers interact.

\subsection{Fisher-information monotone}
\label{sec_I_Fisher}

Focusing on $\varphi = 0, \pi$,
we upper-bounded the isomerization yield $\rho_-(\pi)$
(Sec.~\ref{sec_Thermomaj_Bound}).
But we might wish to calculate $\rho_- (\pi)$,
using resource-theory tools.
We must model the full rotation, $\varphi \in [0, \pi]$,
within the resource theory.
We do so in App.~\ref{sec:LZ_RT}, treating $\varphi$ as a quantum clock~\cite{Pauli_33_Handbuch,Pauli_58_Encyclopedia,Feynman_86_Quantum,Margolus_86_Quantum,Aberg_14_Catalytic,Malabarba_15_Clock,Frenzel_16_Quasi,Woods_16_Autonomous,Erker_17_Autonomous,Woods_18_Quantum}.
The chemical groups' angular momentum, $\ell_\varphi$,
governs the speed at which the clock ``hand'' rotates.
To illustrate how the momentum can affect the dynamics, 
we linearize $H_\elec (\varphi)$ [Eq.~\eqref{eq:H_Elec_Hahn}]
near the avoiding crossing point. 
For simplicity, we assume that the momentum remains constant.

The resulting Hamiltonian has the form of a Landau-Zener model,
\begin{equation}
\label{Eq:HLZ}
   H_\LZ(t)  
   \approx  -v t  \,  ( \ketbra{ \psi_1}{\psi_1}  
   -  \ketbra{ \psi_0}{\psi_0} )  
   +  \frac{\lambda}{2} \, ( \ketbra{ \psi_0}{\psi_1}  +  \hc )
\end{equation}
illustrated near the crossing point in Fig.~\ref{Fig:1}.
The Hamiltonian changes at a speed 
$v\propto | d\varphi/dt |$
that has dimensions of energy/time.
The time, $t$, runs from $-\infty$ to $\infty$
in the Landau-Zener model.

We can understand the model by evaluating limits in an example.
Suppose that the molecule begins in 
the upper diabatic level, $\ket{\psi_1}$.
If $v  \ll  \lambda^2 / \hbar$, 
the state evolves adiabatically,
changing from $\ket{\psi_1}$
but remaining in the upper instantaneous eigenstate.
If $v \gg \lambda^2 / \hbar$, 
the state evolves diabatically,
remaining (approximately) $\ket{\psi_1}$,
which becomes approximately the lower energy eigenstate.
Isomerization in the presence of a thermal bath amounts to 
a dissipative Landau-Zener transition~\cite{ao1989influence,saito2007dissipative,arceci2017dissipative,malla2017suppression}.
We model such transitions within the resource theory in App.~\ref{sec_Dissip_LZ}.
Here, we quantify the postisomerization coherence
in the electronic state
with the quantum Fisher information $I_\Fisher$ 
relative to the Hamiltonian, 
a resource-theory monotone~\cite{Kwon_18_Clock}.

%

$I_\Fisher$ quantifies mixed and pure states' coherences~\cite{Feng_17_Quantifying,Kwon_18_Clock}.\footnote{
$I_\Fisher$ is one of multiple monotones that quantify coherence.
Photoisomers' coherence has been measured, for example, with
rate-constant behavior~\cite{Ashkenazi_99_Photoexcited}.}
Let $\rho$ denote a quantum state
that eigendecomposes as
$\rho =  \sum_j  r_j  \ketbra{ j }{ j }$.
The Fisher information with respect to a Hamiltonian $H$ is 
\begin{align}
   I_\Fisher ( \rho, H )
   = 2 \sum_{j, k}
   \frac{ ( r_j - r_k )^2 }{ r_j + r_k }  \:
   | \bra{ j } H \ket{ k } |^2  \, .
\end{align}
$I_\Fisher$ quantifies the state's ability
to distinguish instants as a quantum clock~\cite{Kwon_18_Clock}.
When evaluated on a pure state, 
$I_\Fisher$ reduces to four times the energy variance,
$\expval{ H^2 }  -  \expval{H}^2$.

We can calculate explicitly the Fisher information 
in the post-photoisomerization state.
We temporarily assume that 
the molecule ends in the \emph{trans} configuration 
(with $\varphi = \pi$),
rather than in a statistical mixture of \emph{cis} and \emph{trans} states.
This assumption simplifies the calculation, which can be generalized,
and is physically realistic.
We eigendecompose the post-photoisomerization state as
$\rho_\final 
=  \rho_\final^\elec  \otimes  
\ketbra{ \varphi {=} \pi }{ \varphi {=} \pi }
=  \sum_{i,j= 0,1}  \rho_{ij} 
\ketbra{\psi_i}{\psi_j}  \otimes  
\ketbra{ \varphi {=} \pi }{ \varphi {=} \pi }$.
For the Landau-Zener Hamiltonian, 
\begin{align}
   \label{eq_I_F_LZ}
    I_\Fisher \LParen \rho_\final^\elec ,  H_\LZ(t_\final)  \RParen
    =  \lambda^2  \left\lvert 1  -  2 \rho_{00} - 4 \, 
    \frac{vt_\final}{\lambda} \, {\rm Re} ( \rho_{01} ) \right\rvert^2 \, ,
\end{align}
wherein the density matrix and 
the Hamiltonian are evaluated at $t=t_\final \gg \lambda/v$. 
The gap, $\lambda$, sets the distance tuned through 
in energy space, due to 
(i) the Hamiltonian's linearization and 
(ii) the order-one change in the angle, $\pi$. 
We have invoked the state's normalization,
$\rho_{00}+\rho_{11}=1$. 
We simplify and bound Eq.~\eqref{eq_I_F_LZ} 
in App.~\ref{sec_I_F_Methods}.
The result is
\begin{align}
   \label{eq_I_F_Plus}
   & I_\Fisher \LParen  \rho_\final^\elec, 
                                 H_\LZ  (t_\final )   \RParen 
   \\ \nonumber &
   \leq  I_\Fisher^{\mathrm{+}} 
   := 16 v^2 t_\final^2 
   e^{- \pi \lambda^2/ (2\hbar v)}
   \left( 1  -  e^{- \pi \lambda^2 / (2\hbar v) }  \right) .
\end{align}
The proportionality to the squared momentum,
$(v t_\final )^2$, quantifies how
underdamping near the avoided crossing
generates more coherence than overdamping would generate.
An expression similar to Eq.~\ref{eq_I_F_Plus} has been derived in the context of parameter estimation on a closed Landau-Zener transition~\cite{yang2017quantum}.

\subsection{Dissipative Landau-Zener transition}
\label{sec_Dissip_LZ}

To study the bath's effects on 
the Landau-Zener evolution of the electronic coherences,
we have evaluated $I_\Fisher$ on the molecule's postrotation state,
$\rho_\final$, again using the Lindblad master equation.
The system Hamiltonian $H$ approximately has 
the Landau-Zener form $H_\LZ(t)$. 
We suppose that the system couples to the bath through the operator
$B  =  \ketbra{ \psi_1}{\psi_1}  -  \ketbra{ \psi_0}{\psi_0}$.
Relative to the $H_\LZ (t)$ eigenbasis near the avoided crossing,
$B$ is represented by a nondiagonal matrix.
Hence $B$ transfers population between energy levels,
at a rate $\Gamma$.
 At frequencies resonant with the photoisomer away from the avoided crossing, the bath's density of states is expected to be vanishingly small. Thus, away from the avoided crossing, the bath's action can be ignored to first order in the system-bath coupling. 
This Lindbladian and alternatives have been ascribed to the Landau-Zener transition elsewhere~\cite{arceci2017dissipative}.
%

To obtain $\rho_\final$, we prepared
the electronic state $\ket{\psi_1}$ 
and simulated evolution from times $t = -t_\final$ to 
$t_\final = 50 \hbar/\lambda$.
For simplicity, we adopt a unit system in which $\lambda=\hbar=1$.
We focus on the dependences on $v$ and $\Gamma$.

\begin{figure}[t]
\includegraphics[width=8.5cm]{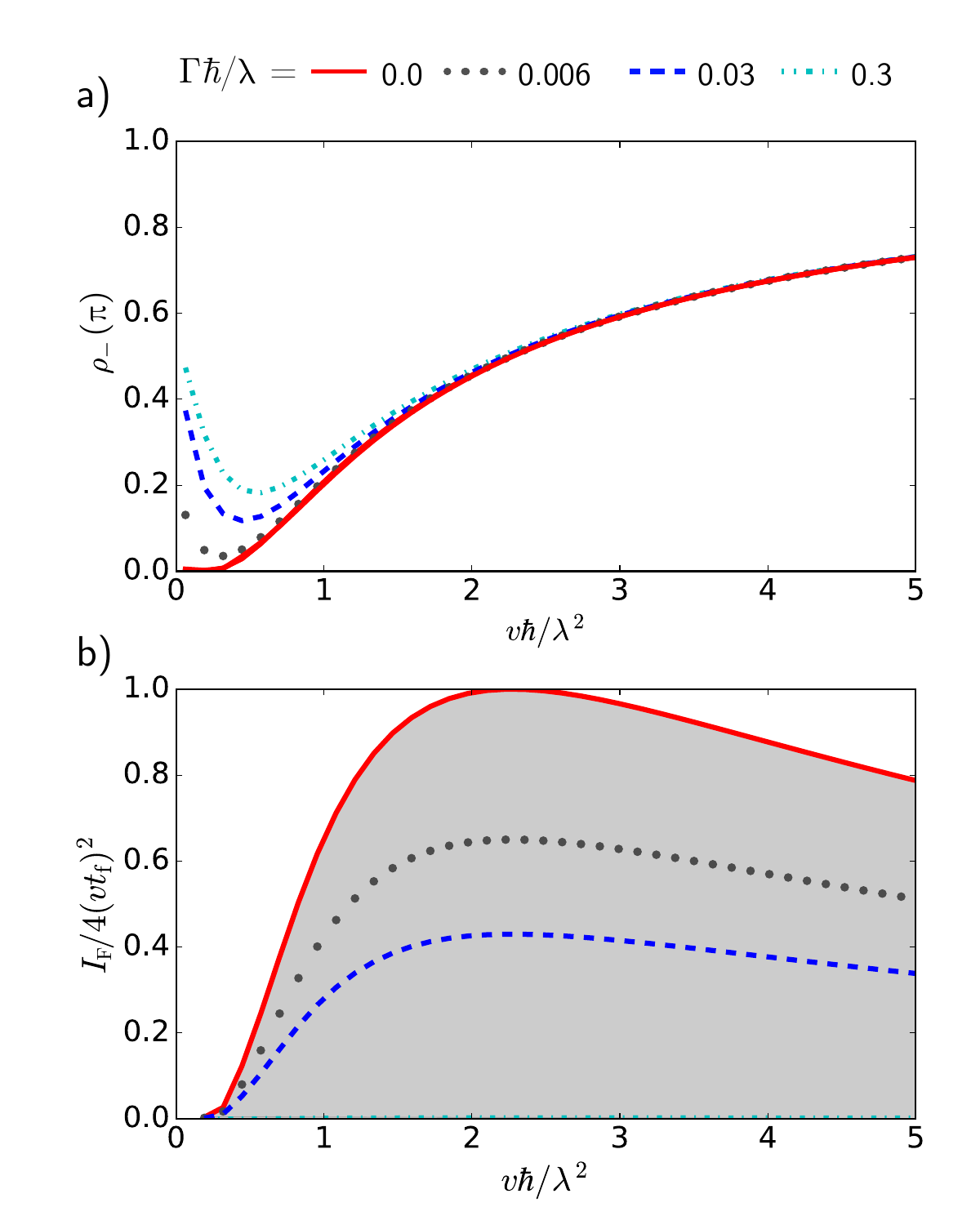}
\label{fig:I_Fisher}
\caption{\caphead{Dissipative Landau-Zener model for 
the lower \emph{trans} state's population 
and for coherence, relative to the energy eigenbasis,
in the electronic state}. 
The initial state, $\ket{ \psi_1 }$, was evolved from
an initial time of $-t_\final$ to $t_\final = 50\, \hbar /\lambda$.
a) Final lower-\emph{trans}-level population as 
a function of transition speed, $v$, 
for different dephasing rates $\Gamma$. 
b) Reduced Fisher information,  
as a function of $v$
for different dephasing rates $\Gamma$. 
The grayed area represents 
the monotone bound~\eqref{eq_I_F_Plus}.
}
\label{fig:Post_LZ}
\end{figure}

Figure~\ref{fig:Post_LZ}a) shows 
the photoisomerization yields, under Landau-Zener dynamics,
for different decohence rates. 
If $\Gamma=0$,  the yield is well-described by 
the canonical Landau-Zener transition probability 
$\rho_{-}(\pi) = \rho_{11} 
\approx \exp \left( - \frac{ \pi \lambda^2 }{2 \hbar v } \right)$. 
At low speeds, $v \hbar /\lambda^2  \ll  1$, 
the yield is small.
The system evolves adiabatically, 
ending in the diabatic state $\ket{\psi_0}$. 
At high speeds, $v \hbar /\lambda^2 \gg 1$,
the yield is greater:
The system lacks time to transition to $\ket{ \psi_0 }$ 
and so remains in $\ket{ \psi_1}$.

Consider raising the phase-damping rate $\Gamma$
at a fixed $v$.
If the speed is low, $v \hbar /\lambda^2 \ll 1$, the yield rises.
If the speed is large, $v /\lambda \gg \Gamma$,
the yield about equals its decoherence-free value,
regardless of $\Gamma$.
The yield minimizes when $\Gamma \approx v/\lambda$:
The decoherence's mixing of energy eigenstates, 
which transfers about half the state's weight to the lower energy level,
balances adiabaticity's preservation of the upper level's weight.

Similar behavior was observed in~\cite{arceci2017dissipative}.
See~\cite{Shimshoni_92_Onset,Whitney_11_Temperature}
for further treatments of coherence in dissipative Landau-Zener transitions.
Whereas earlier work focused mostly on the populations,
we quantify how the electronic energy coherences evolve 
in the dissipative Landau-Zener problem.

Figure~\ref{fig:Post_LZ}b) shows 
the Fisher information scaled by $1/(4 v^2 t_\final^2)$. 
When $\Gamma=0$, $I_\Fisher$ 
adheres to the asymptotic prediction
$I_\Fisher^{\mathrm{+}}$
[Eq.~\eqref{eq_I_F_Plus}],
represented by the grayed region.
The asymptotic bound~\eqref{eq_I_F_Plus}
limits the coherence, 
we verified, for finite $\Gamma$.
Raising $\Gamma$ above 0
decreases the scaled Fisher information toward 0.


The scaled coherence peaks at an intermediate speed
given by Eq.~\eqref{eq_I_F_Plus}.
At this $v$, half the population transitions
from the initial excited state, $\ket{\psi_1}$, 
to the final ground state, $\ket{\psi_1}$.
Transitioning half the population
spreads probability weight evenly 
across the energy levels.
Even spreading accompanies maximal coherence.
This observation agrees with the quantum adiabatic theorem:
If $H_\LZ(t)$ changes slowly, 
the electronic DOF remains in an instantaneous energy eigenstate.
The final state, $\ket{ \Energy_+(\pi) }$, therefore lacks coherence.
If $H_\LZ(t)$ changes quickly,
the state has no time evolve away from $\ket{\psi_1}$.
Since $\ket{\psi_1}$ becomes the $H_\LZ( t_\final )$ ground level,
the final state again lacks coherence.
Hence low and high $v$'s lead to small coherences
that we have quantified with
$I_\Fisher / (4 v^2 t_\final^2)$.

In summary, isomerization partially trades off with 
electronic coherences relative to the energy eigenbasis.
Little population transfer, which is undesirable,
accompanies little coherence.
Little coherence accompanies also
a desirable large population transfer.
Midsize population transfer 
accompanies large coherences.
We have quantified these trends with the Fisher information.
Moreover, electronic energy coherences 
do not straightforwardly promote isomerization in this minimal model.

\section{Discussion}
\label{sec_Outlook}

We have derived fundamental limitations on photoisomerization,
using thermodynamic resource theories.
The bounds are simple, general, 
and derived from few assumptions.
Yet the results shed light on the roles played 
by information, energy, and coherence
in molecules prevalent in natural and artificial materials.

Similar insights may follow from modeling other chemical systems
with thermodynamic resource theories.
Candidates include chlorophyll~\cite{Engel_07_Evidence,Sarovar_10_Quantum,Duan_17_Nature,Chan_18_Single} and photovoltaics~\cite{Jailaubekov_12_Hot,Akselrod_14_Subdiffusive,gilmore2017charge,lin2018thermochromic,Strubbe_18_Thermodynamic}.
 Exciton transport there may be bounded 
as isomerization is here.
Additionally, azobenzene has been attached 
to carbon nanotubes~\cite{Kucharski_14_Templated}.
The attachment raised the isomers' 
capacity for storing solar fuel
by 200\%, though 30\% was expected.
The improvements achievable---and
the engineering effort exerted---might be upper-bounded with 
a variation on our model.

This work leverages resource theories,
which have remained largely abstract,
to solve known problems in experimental systems.
A bridge for thermodynamic resource theories
from mathematical physics to the real physical world
was called for recently~\cite{NYH_17_Toward};
construction has just begun~\cite{Lostaglio_18_Elementary,Lorch_18_Optimal,Holmes_18_Coherent,Alhambra_18_Heat}.
Experimental proposals designed to realize resource-theory results
have provided a valuable first step.
The present paper progresses from artifice
to explaining diverse phenomena 
realized already in nature and in experiments,
to answering questions already asked 
in atomic, molecular, and optical physics and chemistry.
This program may be advanced through
this paper's resource-theory model for Landau-Zener transitions,
which occur across chemistry and many-body physics.

%
%
\begin{acknowledgments}
NYH thanks Bassam Helou, David Jennings, Christopher Perry, and Mischa Woods for conversations.
NYH is grateful for funding from the Institute for Quantum Information and Matter, an NSF Physics Frontiers Center (NSF Grant PHY-1125565) with support from 
the Gordon and Betty Moore Foundation (GBMF-2644),
for a Barbara Groce Graduate Fellowship,
for a KITP Graduate Fellowship
(the KITP receives support from the NSF 
under Grant No. NSF PHY-1125915),
and for an NSF grant for 
the Institute for Theoretical Atomic, Molecular, and Optical Physics 
at Harvard University and the Smithsonian Astrophysical Observatory.
DTL  was supported by UC Berkeley College of Chemistry and by the Kavli Energy NanoSciences Institute.
\end{acknowledgments}

\begin{appendices}

\onecolumngrid

\renewcommand{\thesection}{\Alph{section}}
\renewcommand{\thesubsection}{\Alph{section} \arabic{subsection}}
\renewcommand{\thesubsubsection}{\Alph{section} \arabic{subsection} \roman{subsubsection}}

\makeatletter\@addtoreset{equation}{section}
\def\theequation{\thesection\arabic{equation}}

\section{Every composition of thermal operations is a thermal operation.}
\label{sec_Compose_Th_Ops}

Let $(\rho, H)$ represent a system $\Sys$ of interest.
$\Sys$ evolves under an arbitrary thermal operation $\TO$ as
\begin{align}
   ( \rho, H )  
   \mapsto  \TO ( \rho, H )
   =  \left(  \Tr_\anc  \left( U  \left[  \rho  \otimes  e^{ - \beta H_\Bath } / Z  \right]
                                       U^\dag  \right) ,
                 H  +  H_\Bath  -  H_\anc  \right) .
\end{align}
$H_\Bath$ denotes the bath Hamiltonian, and
$Z  :=  \Tr  \left(   e^{ - \beta H_\Bath } / Z  \right)$
denotes the partition function.
$\anc$ denotes a discarded ancilla
that is not coupled to the rest of the system by any term in
$H + H_\Bath$.
The unitary $U$ preserves the total Hamiltonian:
$[U,  H + H_\Bath]  =  0$.
Let $\tilde{\TO}$ denote another thermal operation:
\begin{align}
   (\rho, H)
   \mapsto  \tilde{\TO} ( \rho,  H )
   =  \left(  \Tr_{\tilde{\anc}}  \left(  \tilde{U}
   \left[  \rho  \otimes  e^{ - \beta H_{ \tilde{\Bath} } }  /  \tilde{Z}  \right]
   \tilde{U}^\dag  \right) ,
   H  +  H_{ \tilde{\Bath} }  -  H_{ \tilde{\anc} }  \right) .
\end{align}
The tilded quantities are defined analogously to their tilde-free analogues.
Consider composing the thermal operations:
\begin{align}
   (\rho, H)
   & \mapsto  \tilde{\TO}  \LParen  \TO  ( \rho, H )  \RParen \\
   \label{eq_Compose}
   & =  \Bigg(  \Tr_{\tilde{\anc}}  \left(  \tilde{U}
   \left\{  \Tr_\anc \left(  U  \left[  \rho  \otimes  
                                                  e^{ - \beta H_\Bath }  /  Z  \right]  U^\dag  \right)
            \otimes  e^{ - \beta H_{ \tilde{\Bath} } }  /  \tilde{Z}  \right\}
   \tilde{U}^\dag  \right) ,
   \\ \nonumber & \qquad \quad
   H  +  H_\Bath  -  H_\anc  +  H_{ \tilde{\Bath} }  -  H_{ \tilde{\anc} }  \Bigg) .
\end{align}
We will show that the composite is a thermal operation:
\begin{align}
   \label{eq_To_Prove}
   \tilde{\TO}  \LParen  \TO  ( \rho, H )  \RParen
   =  \left(  \Tr_{ \bar{\anc} }  \left(  \bar{U}
   \left[  \rho  \otimes  e^{ - \beta H_{ \bar{\Bath} } }  /  \bar{Z}  \right]
   \bar{U}^\dag  \right) , 
   H  +  H_{ \bar{\Bath} }  -  H_{ \bar{\anc} }  \right) ,
\end{align}
for some bath Hamiltonian $\bar{\Bath}$, 
some unitary $\bar{U}$, 
and some uncoupled ancilla $\bar{\anc}$.

Let us return to Eq.~\eqref{eq_Compose}.
The tensoring on of $e^{ - \beta H_{ \tilde{\Bath} } } / \tilde{Z}$
does not change the state of $\anc$.
Nor does $\tilde{U}$ act on the Hilbert space of $\anc$.
Hence the tensoring on and the $\tilde{U}$
commute with the $\Tr_\anc ( . )$.
During the commutation, an $\id_\Bath$ is tensored onto the $\tilde{U}$,
such that $\tilde{U} \otimes \id_\Bath$ acts on
the Hilbert space of $\Sys + \Bath + \tilde{\Bath}$.
The traces compose as 
$\Tr_{ \tilde{\anc} } \LParen \Tr_\anc ( . ) \RParen
=  \Tr_{ \tilde{\anc} + \anc } ( . )
\equiv  \Tr_{ \bar{\anc} } ( . )$:
\begin{align}
   \tilde{\TO}  \LParen  \TO  ( \rho, H )  \RParen 
   \label{eq_Compose2}
   & =  \Bigg(  \Tr_{\bar{\anc}}  \left(  
   \left[  \tilde{U}  \otimes  \id_\Bath   \right]
   \left\{   U   \left[  \rho  \otimes  
                       e^{ - \beta H_\Bath }  /  Z  \right]  U^\dag 
            \otimes  e^{ - \beta H_{ \tilde{\Bath} } }  /  \tilde{Z}  \right\}
   \left[  \tilde{U}  \otimes  \id_\Bath   \right]^\dag   
   \right) ,
   \nonumber \\ & \qquad \quad
   H  +  H_\Bath  +  H_{ \tilde{\Bath} }  -  H_{ \bar{\anc} }  \Bigg) .
\end{align}

The $U$ does not act on the $\tilde{\Bath}$ Hilbert space.
Hence the $U$ commutes with
the tensoring on of $e^{ - \beta H_{ \tilde{\Bath} } }  /  \tilde{Z}$.
In commuting the $U$ out,
we must tensor onto it a $\id_{ \tilde{\Bath} }$.
The unitaries compose as
$(  \tilde{U}  \otimes  \id_\Bath   )
\left(  U  \otimes  \id_{ \tilde{\Bath} }  \right)
=:  \bar{U}$.
This operator is unitary and conserves the global Hamiltonian,
$H + H_\Bath + H_{ \tilde{\Bath} }$. 
Hence
\begin{align}
   \tilde{\TO}  \LParen  \TO  ( \rho, H )  \RParen 
   \label{eq_Compose3}
   & =  \left(  \Tr_{ \bar{\Bath} }  \left(  \bar{U}
   \left[  \rho  \otimes  e^{ - \beta H_\Bath } / Z
                    \otimes  e^{ - \beta H_{ \tilde{\Bath} } }  /  \tilde{Z}
   \right]  \bar{U}^\dag   \right) ,
   H  +  H_\Bath  +  H_{ \tilde{\Bath} }  -  H_{ \bar{\anc} }  \right) .
\end{align}
The composition of two Gibbs states is a Gibbs state:
$e^{ - \beta H_\Bath } / Z
  \otimes  e^{ - \beta H_{ \tilde{\Bath} } }  /  \tilde{Z}
=  e^{ - \beta H_{ \bar{\Bath} } } / \bar{Z}$,
wherein $H_{ \bar{\Bath} }  :=  H_\Bath  +  H_{ \tilde{\Bath} }$.
Equation~\eqref{eq_Compose3} therefore has  the form of Eq.~\eqref{eq_To_Prove}.

\section{Why modes of coherence  transform independently under thermal operations}
\label{sec_Review_Coher_Indep}


This note reviews why modes of coherence 
transform independently under thermal operations.
The independence rests on 
thermal operations' covariance with respect to 
the time translations generated by the system-of-interest Hamiltonian.
The note relies on~\cite{Marvian_14_Modes},
by Marvian and Spekkens,
and on~\cite{Lostaglio_15_Description},
by Lostaglio \emph{et al}.
Marvian and Spekkens  reason generally about U(1) symmetry.
Lostaglio \emph{et al.} tailor Marvian and Spekkens's results
to time-translation symmetry,
basing their conclusions on Marvian and Spekkens's arguments.


\textbf{System of interest, decompositions, and definitions:}
Let $\Sys = (\rho, H)$ denote the initial system.
The operators are defined on a Hilbert space $\Hil$.
Lostaglio \emph{et al.} focus on nondegenerate Hamiltonians $H$.
(The molecule's Hamiltonian lacks degeneracies in our study.)
The Hamiltonian eigendecomposes as
$H \ket{ n }  =  \omega_n  \ket{n}$.

We can decompose the state in terms of the energy eigenbasis:
$\rho   =  \sum_{n, m}  \rho_{nm}  \ketbra{n}{m}$.
The coefficients $\rho_{nm}$ form a matrix.
Column $n$ corresponds to the energy $\omega_n$,
and row $m$ corresponds to the energy $\omega_m$.
The gap $\omega_n - \omega_m$ defines a mode.
Multiple elements $\rho_{nm}$ can correspond to the same mode. 
Summing these elements yields 
``the component of $\rho$ in mode $\omega$'':
\begin{align}
   \rho^\omegaP
   :=  \sum_{n, m \, :  \, \omega_n - \omega_m  =  \omega}
   \rho_{nm}  \ketbra{n}{m} .
\end{align}
The state decomposes in terms of modes as
\begin{align}
   \label{eq_Rho_Mode}
   \rho  =  \sum_\omega  \rho^\omegaP  .
\end{align}
Consider evolving a $\rho^\omegaP$ under the system's Hamiltonian.
The mode acquires a phase:
\begin{align}
   \label{eq_Phase}
   \rho^\omegaP
   & =  e^{ - i H t }  \rho^\omegaP  e^{i H t}
   =  e^{ - i \omega t }  \rho^\omegaP .
\end{align}

\textbf{Thermal operations:}
Let $\TO$ denote any thermal operation in which 
the bath, and only the bath, is discarded:
\begin{align}
   \TO \LParen (\rho, H ) \RParen
   =  (\sigma, H) ,
\end{align}
for some density operator $\sigma$ defined on $\Hil$.
The thermal operation is covariant with respect to
the translations generated by $H$:
Let $H_\Bath$ denote the Hamiltonian of the bath system
that couples to $\Sys$,
$\Bath  =  ( e^{ - \beta H_\Bath } / Z,  H_\Bath )$. 
For every operator $O$ defined on $\Hil$,
\begin{align}
   \label{eq_Trans_Help1}
   \TO \left(  e^{ - i H t }  O  e^{i H t}  \right)
   & =  \Tr_\Bath  \left( U  \left[
   e^{ - i H t }  O  e^{i H t}  \otimes
   e^{ - \beta H_\Bath } / Z  
   \right]  U^\dag  \right) \\
   \label{eq_Trans_Help2}
   & =  \Tr_\Bath  \left( U  \left[
   e^{ - i H t }  O  e^{i H t}  \otimes
   e^{ - i H_\Bath t }  ( e^{ - \beta H_\Bath } / Z )  e^{i H_\Bath t}
   \right]  U^\dag  \right) \\
   \label{eq_Trans_Help3}
   & =  \Tr_\Bath  \left( U  \left\{
   e^{ - i (H + H_\Bath) t }
   \left[  O  \otimes  e^{ - \beta H_\Bath } / Z  \right]
   e^{ i (H + H_\Bath) t }
   \right\}  \right)  \\
   \label{eq_Trans_Help4}
   & =  \Tr_\Bath  \left(
   e^{ - i (H + H_\Bath) t }  U
   \left[  O  \otimes  e^{ - \beta H_\Bath } / Z  \right]
   U^\dag  e^{ i (H + H_\Bath) t }  \right)  \\
   \label{eq_Trans_Help5}
   & =  e^{ - i H t }   \Tr_\Bath  \left(
   U  \left[  O  \otimes  e^{ - \beta H_\Bath } / Z  \right]
   U^\dag  \right)  e^{ i H t }  \\
   \label{eq_Trans_Help6}
   & =  e^{ - i H t }   \TO ( O )  e^{ i H t } .
\end{align}
The first equality follows from the thermal operation's definition.
Equation~\eqref{eq_Trans_Help2} follows from 
the bath state's invariance under 
the time translations generated by $H_\Bath$.
Equation~\eqref{eq_Trans_Help3} follows from 
the commutation of $H$ with $H_\Bath$.
The commutation of $U$ with $H + H_\Bath$,
codified in the definition of $\TO$,
implies Eq.~\eqref{eq_Trans_Help4}.
Equation~\eqref{eq_Trans_Help5} represents
the equivalence of two protocols:
(i) evolving the whole system under $H + H_\Bath$,
then discarding the bath, and
(ii) discarding the bath, then evolving $\Sys$ under $H$.

\textbf{Claim:}
Let 
\begin{align}
   \label{eq_Sigma_Decomp}
   \sigma  =  \sum_\omega  \sigma^\omegaP
\end{align}
denote the mode decomposition of $\sigma$.
We wish to prove that the coherences relative to the $H$ eigenbasis
transform independently under 
the thermal operations in which the bath, and only the bath, is traced out:
\begin{align}
   \label{eq_Claim}
   \TO \left(  \rho^\omegaP  \right)
   =  \sigma^\omegaP  .
\end{align}
We substitute $O = \rho^\omega$ into Eq.~\eqref{eq_Trans_Help6}:
\begin{align}
   e^{ -iHt }  \TO \left(  \rho^\omegaP  \right)
   e^{ iHt}
   \label{eq_Final_Help1}
   & =  \TO  \left(  e^{ -iHt}  \rho^\omegaP  e^{iHt}  \right)  \\ 
   \label{eq_Final_Help2}
   & =  \TO  \left(  e^{-i \omega }  \rho^\omegaP  \right)  \\ 
   \label{eq_Final_Help3}
   & =  e^{-i \omega }  
   \TO  \left(  \rho^\omegaP  \right) .
\end{align}
Equation~\eqref{eq_Final_Help2} follows from Eq.~\eqref{eq_Phase}.
Equation~\eqref{eq_Final_Help3} follows from
the convex linearity of quantum operations, 
which include thermal operations.\footnote{
Convex-linear operators $O$ map 
convex combinations to convex combinations:
If $\Set{ p_j }$ denotes a probability distribution
and $\rho_j$ denotes a density operator defined on $\Hil$,
$O \left( \sum_j p_j  \rho_j  \right)
=  \sum_j  p_j  O (\rho_j)$.}
The left-hand side of Eq.~\eqref{eq_Final_Help1},
with Eq.~\eqref{eq_Final_Help3}, implies that
$\TO \left(  \rho^\omegaP  \right)$ belongs to mode $\omega$.
Therefore, by Eq.~\eqref{eq_Sigma_Decomp},
$\TO \left(  \rho^\omegaP  \right)$ is $\sigma^\omegaP$.

\section{Details about Lindblad evolution}
\label{sec_KineticMethods}

We choose Lindblad operators  
$B_{\Energy_\mu(\varphi),\Energy_{\mu '}(\varphi')}
= \ketbra{ \Energy_\mu (\varphi) }{ \Energy_{\mu'} (\varphi')}$,  
for each pair of energy eigenstates.
Each $B_i$ dissipates at a rate  $\Gamma_i$
assumed to satisfy local detailed balance,
\begin{align}
   \frac{\Gamma_{\Energy_\mu(\varphi),\Energy_{\mu '}(\varphi')}}{\Gamma_{\Energy_{\mu '}(\varphi'),\Energy_\mu(\varphi)}} 
   = e^{- \beta [ \Energy_\mu(\varphi)-\Energy_{\mu '}(\varphi') ]} ,
\end{align}
so the system relaxes toward a thermal state. 
To model the slowness of thermal isomerization, we set 
$\Gamma_{\Energy_-(0),\Energy_{-}(\pi)} 
= \Gamma_{\Energy_{-}(\pi),\Energy_-(0)} =0$. 
Also processes involving the high-energy state 
$\ket{ \Energy_+(\pi) }$ can be ignored. The numerical evaluations of the master equation were accomplished using QuTip code~\cite{johansson2013qutip}.

%

The Lindblad equation
can be solved analytically, 
though the solution is complicated.
We can gain intuition from the regime
\begin{align}
   \label{eq_Gamma_Condn}
   \Gamma_{\Energy_+(0),\Energy_-(\pi)} 
   \gg \Gamma_{\Energy_+(0),\Energy_-(0)} ,
\end{align}
in which relaxation into the \emph{trans} state 
is kinetically preferred.
Furthermore, satisfying Ineq.~\eqref{eq_Gamma_Condn} and
\begin{align}
   \label{eq_E_Condn}
   E_1\gg \Delta E 
\end{align}
simultaneously enables the Lindblad evolution to saturate
the resource-theory bounds, as we shall see.
Under Ineq.~\eqref{eq_Gamma_Condn}, 
the ground \emph{trans} state has a population of
\begin{align}
   \rho_-(\pi) 
   = \frac{1}{1+e^{\beta (\Delta E - E_1)}} \left ( 1 - e^{- t k} \right ) 
\end{align}
at early times $t \ll 1/ \Gamma_{\Energy_+(0),\Energy_-(0)}$.
The population grows, from 0 at $t = 0$,
with the effective rate 
$k  =  \Gamma_{\Energy_+(0),\Energy_-(\pi)} 
+ \Gamma_{\Energy_-(\pi),\Energy_+(0)}$. 
Our qualitative results are insensitive to the specific parameters studied,
if the constants satisfy
Ineqs.~\eqref{eq_Gamma_Condn} and~\eqref{eq_E_Condn}.
In the intermediate-time limit, 
$1/\Gamma_{\Energy_+(0),\Energy_-(0)} \gg t \gg 1/k$, 
the molecule likely isomerizes:
$\rho_-(\pi)$ approaches 1.
\section{Resource-theory model for dissipative Landau-Zener problem}
\label{sec:LZ_RT}

We model the dissipative Landau-Zener transition of Sec.~\ref{sec_Dissip_LZ}
within the thermodynamic resource theory.
This quantum-information approach may offer insights into
the interplay between energy and coherence
in Landau-Zener transitions across chemistry and condensed matter.

\textbf{Clocks:}
The Landau-Zener problem involves a speed $v$ and so time.
In contrast, every thermal operation $\mathcal{T}$
has time-translation covariance~\cite{Lostaglio_15_Quantum}:
$e^{ - i H t }  \mathcal{T} (\rho)  e^{ i H t }
=  \mathcal{T}  \left(  e^{ - i H t }  \rho  e^{ i H t }  \right)$,
if $\rho$ denotes a state governed by
the Hamiltonian $H$.
We must therefore introduce a clock
into our resource-theory formalism~\cite{Kitaev_04_Superselection,vanEnk_05_Quantifying,Bartlett_07_Reference,Vaccaro_08_Tradeoff,Gour_08_Resource,Aberg_14_Catalytic,Malabarba_15_Clock,Frenzel_16_Quasi,Woods_16_Autonomous,Erker_17_Autonomous,Woods_18_Quantum}.
Clocks have been modeled as instances of more-general reference frames~\cite{Malabarba_15_Clock,Frenzel_16_Quasi,Woods_16_Autonomous,Erker_17_Autonomous,Woods_18_Quantum},
A reference frame is a resource that effectively 
lifts a superselection rule such as energy conservation.
A good clock occupies an even superposition of many energy eigenstates~\cite{Kitaev_04_Superselection,vanEnk_05_Quantifying,Bartlett_07_Reference,Vaccaro_08_Tradeoff,Gour_08_Resource,Aberg_14_Catalytic}
and so has substantial coherence.
A clock can dictate which Hamiltonian governs the system of interest
at any given instant.

\textbf{Molecular clock:}
The molecular configuration $\varphi$ serves as a clock
in Eq.~\eqref{eq:H_Mol}.
The rotating chemical group, shown in Fig.~\ref{Fig:1},
serves as the clock hand.
When $\varphi = 0$, the hand effectively points to 12:00,
and the \emph{cis} Hamiltonian
$H_\elec (\varphi{=}0)$ governs the electronic DOF.
When $\varphi = \pi$, the hand effectively points to 6:00,
and the \emph{trans} Hamiltonian
$H_\elec ( \varphi{=}\pi )$ governs the electronic DOF.

A reliable clock hand has at least two properties:
(i) Which number the hand points to can be distinguished.
(ii) The clock hand sweeps across the clock face steadily.
To serve as a good clock, therefore, 
the chemical group should have
a well-defined angular position $\varphi$ 
and a well-defined angular momentum $\ell_\varphi$.
The chemical group can have both 
due to its semiclassicality:
Being large, the chemical group 
collides with other molecules frequently.
The collisions localize the chemical group spatially.
Being heavy, the chemical group is expected to have 
a large angular momentum:
$\expval{ \ell_\varphi }  \sim  \frac{ \hbar }{ m r }$,
wherein $m$ denotes the mode's effective mass 
and $r$ denotes the molecule's radius.
The configuration occupies a state analogous to
a coherent state of light~\cite{Mandel_95_Optical,Frenzel_16_Quasi}
and to the Gaussian clock state in~\cite{Woods_16_Autonomous}.

\textbf{Evolution:}
As the molecule rotates, the clock hand ticks forward.
To simplify the discussion, 
we will use the Schr\"odinger picture.
In contrast, many thermodynamic-resource-theory calculations 
are performed in the interaction picture:
Consider a system $\Sys$ that interacts with a bath $\Bath$
during a thermal operation $\mathcal{T}$.
A Hamiltonian-conserving unitary $U$
evolves $\Sys + \Bath$:
$[U, H_\Sys + H_\Bath]  =  0$.
The conservation enables us to ignore
the evolution generated by $H_\Sys + H_\Bath$.

We discretize $\varphi  \in  [0, 2\pi]$ into $2f$ values,
for a fixed value of $f$:
\begin{align}
   \label{eq:HMol_Discrete}
   H_\mol  
   =  \sum_{j = 0}^{2f - 1}  \left[
   H_\elec ( \varphi_j )  \otimes
   \ketbra{ \varphi_j }{ \varphi_j }
   +  \id_\elec  \otimes
   \frac{ \ell_\varphi^2 }{ 2 \Moment }  \right]  \, .
\end{align}
When $t = 0$, $\varphi_0 = 0$,
and when $t = t_f$, $\varphi = \pi$.
We extend the angle to be $\varphi  \in  [0, \, 2 \pi)$, such that
$\varphi_{2f}  =  \varphi_0$.
This extension can facilitate the mathematics
regardless of whether a photoisomer opens
to $\varphi = \pi$ or to $\varphi = 2 \pi$.
We have discretized $\varphi$ for physical realism
and for convenience of application of resource-theory results.
Discretizing $\varphi$, using an analogue of 
the discrete-variable representation basis~\cite{Light_85_Generalized},
is formally equivalent to
truncating the energy eigenbasis with a high-energy cutoff.
Because we focus on finite-energy excitations,
instantaneous energy eigenstates
with energies much higher than the initial state's
can be ignored:
They will not couple to the evolution
during the Landau-Zener transition.
Resource theorists often prefer to study discrete systems.
Yet a continuous version of the Lorenz curve,
the cumulous distribution function, exists.
Furthermore, a model for extending resource-theory results to continuous
appears in App.~G of~\cite{Skrzypczyk_13_Extracting}.

We model the evolution as 
a sequence of two alternating time steps:
(i) A ``tick operation'' shifts the clock hand forward,
changing the Hamiltonian $H_\elec (\varphi)$ 
experienced by the electronic DOF.
We model the electronic state as approximately constant
during this time step.
(ii) The new $H_\elec (\varphi)$ evolves the electronic state
for a time $\Delta t$.
The greater the $\Delta t$, 
the more slowly the molecule rotates.

\textbf{Speeds:}
This model has three regimes:
the sudden-quench limit, the quantum-adiabatic limit,
and intermediate speeds.
Let us introduce these regimes in turn.
To facilitate understanding, we suppose that
the electronic DOF begins in an eigenstate $\ket{ \Energy_+ (0) }$
of $H_\elec (0)$.

In the sudden-quench limit, 
$\Delta t  \ll  \frac{ \hbar }{ \Energy_+ (\varphi) - \Energy_- (\varphi) }$
for all $\varphi$.
No intermediate $H_\elec (\varphi)$'s have time
to evolve the electronic DOF.
The electronic state remains $\ket{ \Energy_+ (0) }$,
while $H_\elec (\varphi)$ changes drastically.
The final electronic state may therefore have coherence 
relative to the final energy eigenbasis.

In the quantum-adiabatic limit,
$\Delta t  \gg  \frac{ \hbar }{ \Energy_+ (\varphi) - \Energy_- (\varphi) }$
for all $\varphi$.
After the first time step, $H_\elec (\varphi_1)$
evolves the electronic state $\ket{ \Energy_+ (0) }$.
A matrix diagonal relative to the $H_\elec (\varphi_0)$ eigenbasis
represents the initial state, $\ketbra{ \Energy_+(0) }{ \Energy_+(0) }$,
while an off-diagonal matrix represents $H_\elec (\varphi_1)$.
The off-diagonal elements change the state.
The change is sizable, because $\Delta t$ is large.
In the $\Delta t \to \infty$ limit,
the change evolves the state into
an eigenstate of $H_\elec (\varphi_1)$.

In the intermediate regime,
$\Delta t  \approx  
\frac{ \hbar }{ \Energy_+ (\varphi) - \Energy_- (\varphi) }$.
Each new $H_\elec (\varphi)$ updates the state, 
but not to an eigenstate of the instantaneous Hamiltonian.

\textbf{Ticking operation:}
To introduce the ticking operation,
we temporarily disregard the bath.
The operator
\begin{align}
   \label{eq:U_Tick}
   U_\tick  
   :=  \id_\elec  \otimes
   \sum_{j=0}^{2f - 1}  
   \ketbra{ \varphi_{j+1} }{ \varphi_j } 
\end{align}
rotates the molecule.
The operator is unitary,
$U_\tick^\dag U_\tick  =  U_\tick U_\tick^\dag  
=  \id_\elec  \otimes  \id_\varphi$,
by the modularity of $\varphi$.

The system-and-clock Hamiltonian can generate $U_\tick$
if the clock evolves under a Hamiltonian proportional to its momentum~\cite{Pauli_33_Handbuch,Pauli_58_Encyclopedia,Feynman_86_Quantum,Margolus_86_Quantum,Woods_16_Autonomous,Malabarba_15_Clock}.
This requirement stipulates, in our case, that
$H_\varphi  =  c \ell_\varphi$.
The constant $c  \in  \mathbb{R}$ can be set to one.
This Hamiltonian has an eigenspectrum unbounded from below
and so is unphysical.
If the clock Hamiltonian were physical,
we could use this result to model, resource-theoretically,
a molecule tumbling by itself.

Nevertheless, we can use the ideal clock 
to understand how the molecule's angular DOF
serves as an imperfect clock.
The ideal clock's $H_\varphi  \propto  \ell_\varphi$ 
kicks the clock hand forward:
$H_\varphi$ generates a unitary
$e^{ - \frac{i}{\hbar} \ell_\varphi t }$
that evolves the clock  as
$\ket{ \varphi_j }  
\mapsto  e^{ - \frac{i}{\hbar} \ell_\varphi t }  \ket{\varphi_j}
=  \ket{ \varphi_j + t }
\equiv  \ket{ \varphi_{j+1} }$.
Again, we have translated the notation of~\cite{Malabarba_15_Clock}
into our notation.

Our clock's Hamiltonian equals
the kinetic energy in Eq.~\eqref{eq:H_Mol}:
$H_\varphi = \frac{ ( \ell_\varphi )^2 }{ 2 \Moment }$.
This $H_\varphi$ not only shifts the angular DOF forward,
but also spreads out the DOF's state in position space.
This spreading is not expected to change the state much,
due to the chemical group's near-classicality.
To reconcile the molecule's Hamiltonian
with the quantum-clock formalism more precisely,
one might adapt~\cite{Woods_16_Autonomous}.
In~\cite{Woods_16_Autonomous}, 
Woods \emph{et al.} approximate the 
$H_c  \propto  \ell_\varphi$ clock
with an oscillator whose Hilbert space has
a finite dimensionality.

%
%
%
\textbf{Dissipative ticking operation:}
Let us reincorporate the bath into the model.
While rotating, the molecule jostles bath particles.
They carry off energy dissipated as $H_\elec (\varphi)$ changes.

To model the dissipation, 
we assume that the bath Hamiltonian's spectrum is dense
and contains gaps of all sizes:
For every electronic energy eigenstate 
$\ket{ \Energy_\pm (\varphi_j) }$,
$H_\Bath$ has eigenstates
$\ket{ \EBath_j^\pm }$ and $\ket{ \EBath_{j+1}^\pm }$
such that the energy leaving the molecule enters the bath:
\begin{align}
   \label{eq:Degen_Cond}
   & \bra{  \Energy_\pm (\varphi_j)  }  H_\elec (\varphi_j)
   \ket{  \Energy_\pm (\varphi_j)  }
   +  \bra{  \EBath_j^\pm  }  H_\Bath  \ket{  \EBath_j^\pm  } 
   \\ \nonumber &
   =  \bra{  \Energy_\pm (\varphi_j)  }  
   H_\elec (\varphi_{j+1})
   \ket{  \Energy_\pm (\varphi_j)  }
   +  \bra{  \EBath_{j+1}^\pm  }  
   H_\Bath  
   \ket{  \EBath_{j+1}^\pm  }  .
\end{align}
For simplicity, we assume that $\Bath$ has
only one pair $\ket{ \EBath_j^\pm }$, $\ket{ \EBath_{j+1}^\pm }$
that satisfies condition~\eqref{eq:Degen_Cond}, for each $j$.
This assumption can be relaxed.

The isometry
\begin{align}
   \bar{U}_\tick 
   :=  \sum_{j=0}^{2f - 1}  \sum_{ \mu = \pm }
   \ketbra{  \Energy_\mu ( \varphi_j )  }{  \Energy_\mu (\varphi_j)  }  \otimes
   \ketbra{ \varphi_{j+1} }{ \varphi_j }  \otimes
   \ketbra{  \EBath^\mu_{j+1}  }{  \EBath^\mu_j  }
\end{align}
rotates the molecule while transferring energy 
from the molecule to the bath.
$\bar{U}_\tick$ preserves the average energy by design:
If $\rho$ denotes the initial state of the molecule-and-bath composite,
$\Tr \left[  \rho    ( H_\mol + H_\Bath )  \right]
=  \Tr \left[  
\bar{U}_\tick   \rho   \bar{U}_\tick^\dag 
( H_\mol + H_\Bath )  \right]$.

Three opportunities remain for future work:
(i) $\bar{U}_\tick$ should be elevated
from an isometry to a unitary.
(ii) The energy conservation should be elevated
from average to exact:
$\bar{U}_\tick$ should commute with the global Hamiltonian.
Exact conservation might require
further use of reference frames.
Some thermodynamic-resource-theory works, however,
have required only average energy conservation~\cite{Skrzypczyk_14_Work}.
(iii) The dissipation should be generalized to
arbitrary molecule-bath coupling strengths.
Suppose that the bath occupies the state $\ket{ \EBath^\pm_j }$,
being receptive to energy transfer.
$\bar{U}_\tick$ transfers energy deterministically,
reflecting strong coupling.
But the coupling might be weak in physical reality.
The molecule can have some probability $< 1$ of dissipating
even if the bath is in $\ket{ \EBath^\pm_j }$.
One would adapt the first set-off equation in~\cite[App.~B]{Lostaglio_18_Elementary},
attributed to~\cite{Aberg_13_Truly},
to many-level systems.

%
%
%
\textbf{Sequence of time steps:}
Let $\rho_\mol$ denote the molecule's initial state.
The first two time steps evolve the state as
\begin{align}
   \rho_\mol
   & \mapsto  \Tr_\Bath  \left[
   \bar{U}_\tick
   \left(  \rho_\mol  \otimes   
            \frac{ e^{ - \beta H_\Bath } }{ Z_\Bath }  \right)
   \bar{U}_\tick^\dag  \right]
   =:  \rho'_\mol \\
   & \mapsto  e^{ - i H_\mol  \Delta t }
   \left(  \rho'_\mol  \right)
   e^{ i H_\mol  \Delta t }  \, .
\end{align}
These two evolutions are repeated $f - 1$ times.

\section{Simplification and bounding of the Fisher information}
\label{sec_I_F_Methods}

We derive Ineq.~\eqref{eq_I_F_Plus}
by simplifying and bounding Eq.~\eqref{eq_I_F_LZ}.
Consider the long-time limit, and suppose that $\rho_{01}  \neq  0$.
The term proportional to $v t_\final$ dominates Eq.~\eqref{eq_I_F_LZ}, and
\begin{align}
   I_\Fisher
   \sim  16 v^2 t_\final^2
   | {\rm Re} ( \rho_{01} ) |^2 .
\end{align}
The simplified $I_\Fisher$ depends on the off-diagonal element $\rho_{01}$
of the density matrix relative to the diabatic basis.
Furthermore, this $I_\Fisher$ is proportional to the squared momentum,
$(v t_\final)^2$. 
This proportionality quantifies how
underdamping near the avoided crossing
generates more coherence than overdamping would generate.
The long-time population is well-approximated by 
the Landau-Zener transition probability, 
$\rho_{11} 
\approx \exp  \left( - \frac{ \pi \lambda^2 }{2 \hbar v }  \right)$~\cite{nitzan2006chemical}. 
We approximate the long-time density matrix's off-diagonal elements
by neglecting the phase:
$\rho_{01} 
\approx \exp \left( - \frac{ \pi \lambda^2 }{4 \hbar v}  \right)
\sqrt{1  -  \exp  \left( - \frac{ \pi \lambda^2 }{2\hbar v}  \right)}$. 
Within the approximate treatment of 
the avoided crossing, therefore,
the final-state coherence in the electronic state is upper-bounded by 
\begin{align}
   \label{eq_I_F_Plus_Methods}
   I_\Fisher^{\mathrm{+}} 
   = 16 v^2 t_\final^2 
   e^{- \pi \lambda^2/ (2\hbar v)}
   \left( 1  -  e^{- \pi \lambda^2 / (2\hbar v) }  \right) .
\end{align}

We have bounded a closed system's $I_\Fisher$.
$I_\Fisher$ is a thermal-operations monotone.
Hence any action of the bath results in 
$I_\Fisher \LParen  \rho_\final^\elec, 
                                 H_\LZ  (t_\final )   \RParen
   < I_\Fisher^{\mathrm{+}}$.

\section{Work extraction and injection}
\label{sec_Work}

Using the resource-theory framework,
we show that work can be extracted from postisomerization coherence,
if molecules interact and obey indistinguishable-particle statistics.
We also calculate the minimal work required to photoexcite the molecule.
We use two resource-theory tools:
(i) one-shot information theory, which generalizes 
Shannon information theory to small scales,
and (ii) quantum-thermodynamic results about
extracting work from coherence.

%
%
%
\subsubsection{Extracting work from \\postisomerization coherence}
\label{sec_Extract_W}

The postisomerization state $\rho_\final$ 
can have coherence between unequal-energy electronic levels 
$\ket{ \Energy_\pm (\pi) }  \ket{ \varphi {=} \pi }$.
Work can be extracted from coherence between degenerate levels,
resource-theory results show~\cite{Skrzypczyk_13_Extracting,Skrzypczyk_14_Work,Kwon_18_Clock,Korzekwa_16_Extraction}.
We can generate degenerate-level coherence
from unequal-energy coherence,
using multiple copies of the system.
The work comes from coherence because
the extraction preserves the state's energy diagonal.

Consider two molecules that dissipate weakly
during isomerization.
Having begun in a nearly pure state,
the isomers end nearly in some pure state 
$\ket{ \chi } \ket{ \varphi {=} \pi }^{\otimes 2}$.
The electronic factor has the form
\begin{align}
   \label{eq_W_From_Coh_State}
   \ket{\chi}
& =  \sqrt{ p_{--} }  \ket{ \Energy_-(\pi),  \Energy_-(\pi) }  
+  \sqrt{ p_{-+} }  \ket{ \Energy_-(\pi),  \Energy_+(\pi) } 
 \\ \nonumber & \quad
+  \sqrt{ p_{+-} }  \ket{ \Energy_+(\pi),  \Energy_-(\pi) }  
+  \sqrt{ p_{++} }  \ket{ \Energy_+(\pi),  \Energy_+(\pi) } .
\end{align}

One can initiate work extraction by measuring the system's energy,
e.g., to ascertain how much work is expected 
and so to guide instrument calibration.
Suppose that (i) the equal-energy eigenstates have equal prefactors,
$\sqrt{ p_{-+} }  =  \sqrt{ p_{+-} }$,
and (ii) the greatest Gibbs-rescaled probability is $p_{+-}$:
\begin{align}
   \label{eq_W_From_Coh_Condn}
   \arg \Max_{ \tilde{\Energy} } &
   \Set{  p_{\mu \nu} 
   e^{ \beta \tilde{\Energy} }  }  
   =  \Energy_+ (\pi)  +  \Energy_- (\pi) .
\end{align}
Suppose, further, that the measurement yields the degenerate energy,
$\Energy_- (\pi)  +  \Energy_+ (\pi)$.
The system is projected onto a pure state,
$\frac{1}{ \sqrt{2} } \:
[ \ket{ \Energy_-(\pi),  \Energy_+(\pi) }
 +  \ket{ \Energy_+(\pi),  \Energy_-(\pi) } ]$,
in a two-dimensional space.
The pure state has more informational value than 
the same-energy-diagonal mixed state,
$\frac{1}{2} [ \ketbra{ \Energy_-(\pi) }{ \Energy_-(\pi) }
+  \ketbra{ \Energy_+(\pi) }{ \Energy_+(\pi) } ]$.
The pure state can be decohered to the mixed state
while the extra value is extracted as work
(see Fig.~2 in~\cite{Kwon_18_Clock}).

Let us illustrate how a two-molecule state 
can satisfy the criterion~\eqref{eq_W_From_Coh_Condn}.
Though artificial, the illustration demonstrates achievability.
Suppose that the molecules occupy a small, symmetric structure.
Their real-space wave functions might overlap considerably,
rendering the molecules indistinguishable~\cite{Fisher_18_Quantum}.
The molecules would occupy 
a symmetric or an antisymmetric state,
depending on their total spins~\cite{Fisher_18_Quantum}.
Hence the electronic DOFs could occupy an antisymmetric state~\cite{Fisher_18_Quantum}.
Suppose that a Heisenberg Hamiltonian 
$\propto \bm{\sigma} \cdot \bm{\sigma}$
couples the electronic DOFs during photoisomerization.
Let the Hamiltonian's proportionality constant be positive.
Suppose, further, that the photoisomers thermalize with a $T = 0$ bath.
The electronic state cools to the ground state,
the singlet 
$\frac{1}{ \sqrt{2} } [ \ket{ \Energy_+(\pi),  \Energy_- (\pi) }
-  \ket{ \Energy_- (\pi),  \Energy_+ (\pi) } ]$,
satisfying Eq.~\eqref{eq_W_From_Coh_Condn}.
The photoisomers could decouple quickly after photoisomerizing.
The total Hamiltonian would return to $H_\mol + H_\mol$
while the state remained a singlet.
Work could be extracted from the state's coherence,
if an agent acted quickly enough.
Granted, the decoupling would cost (positive or negative) work.
But that work could, in principle, come from the photoisomers themselves,
or from structures to which the photoisomers were attached,
rather than from the battery charged later from the coherence.

%
%
%
\subsubsection{Minimal work required to photoisomerize}
\label{sec_Min_W}

How much work must one invest to excite a molecule from
$\exp( - \beta H_\mol ) / Z_\mol$ to 
$\ket{ \psi_1 }  \ket{ \varphi{=}0 }$?
One might expect an average of about
$\Energy_+ (0)  -  \Energy_- (0)$.
But the single-photon limit invites us to consider
the minimal work $W_\Min$ required in any one shot.
One-shot work can be calculated in thermodynamic resource theories~\cite{delRio_11_Thermodynamic,Egloff_15_Measure,Horodecki_13_Fundamental,Aberg_13_Truly,Renes_14_Work,NYH_16_Beyond,Faist_18_Thermodynamic}.
Calculations rely on
one-shot information theory~\cite{Renner_05_Security,Datta_09_Min,Berta_13_Quantum,Leditzky_16_Relative,Wilde_17_Quantum},
which extends Shannon information theory~\cite{Shannon_48_Mathematical}
to small scales and few trials.

Work can be defined in terms of a battery~\cite{Horodecki_13_Fundamental}.
The battery can manifest as an oscillator
governed by a Hamiltonian
like
$H_\Laser$ ~\cite{Skrzypczyk_13_Extracting,NYH_16_Beyond,NYH_18_Beyond,Aberg_14_Catalytic}.
A battery performs work while facilitating
a system-of-interest transformation
from some state $\rho$ to some state $\rho'$.
The work is positive, $\hbar \omega > 0$, if the battery de-excites:
$(\rho \otimes \ketbra{ n_\omega }{ n_\omega } )
\mapsto
(\rho'  \otimes  \ketbra{ n_\omega - 1 }{ n_\omega - 1 })$.
We regard the light source as consisting of 
batteries of various gaps $\hbar \omega$
(see Sec.~\ref{sec_RT_Model_Isomer}).

Consider creating one copy of 
an arbitrary energy-diagonal system
$\LParen \Decohere(\rho), H \RParen$ 
from a thermal system
$\left( \exp[ - \beta H ] / Z,  H  \right)$.
The minimal work required has been shown to equal
\begin{align}
   \label{eq:W_Min_Gen}
   W_\Min \LParen \Decohere(\rho), H \RParen
   =  \frac{1}{\beta}
   D_\Max \left(  \rho  ||  e^{ - \beta H } / Z  \right)  
\end{align}
\cite{Horodecki_13_Fundamental}.
The max relative entropy between 
quantum states $\rho$ and $\sigma$ is defined as
\begin{align}
   D_\Max ( \rho || \sigma )
   :=  \log  (  \Min \Set{  
   c  \in  \mathbb{R}  :  \rho  \leq  c  \sigma  }  )  \, .
\end{align}
We set logarithms to be base-$e$ in this paper.
$D_\Max$ is well-defined if 
the first state's support lies in the second state's:
${\rm supp} (\rho)  \subseteq  {\rm supp} (\sigma)$.
This entropy quantifies how well $\rho$ and $\sigma$ can be distinguished
in a particular trial of state discrimination~\cite{Garner_18_One}.

Using these results, we calculate the work required 
to excite the molecule's thermal state to
$\ket{\psi_1}  \ket{\varphi{=}0}$.
We notate the energy diagonal as
\begin{align}
   & \Decohere ( \ketbra{\psi_1}{\psi_1}  \otimes 
   \ketbra{\varphi{=}0}{\varphi{=}0} )
   \nonumber \\ & 
   =  \sum_{\mu = \pm}  p_\mu
   \ketbra{ \Energy_\mu (0) }{ \Energy_\mu (0) }
   \otimes \ketbra{\varphi{=}0}{\varphi{=}0}  \, .
\end{align}
Substituting into Eq.~\eqref{eq:W_Min_Gen} yields
\begin{align}
   \label{eq_W_Min_Gen_Result}
   & W_\Min \LParen 
   \Decohere(\ketbra{\psi_1}{\psi_1}  
                     \otimes  \ketbra{\varphi{=}0}{\varphi{=}0} ),  
   H_\mol  \RParen
   \nonumber \\ & 
   =  \max_{ \mu = \pm } \Set{ 
   \Energy_\mu (0) 
   - \frac{1}{\beta}  \log \left( 1 / p_\mu \right)  }
   -  \left(  - \frac{1}{\beta}  \log  Z_\mol  \right)  \, .
\end{align}

The expression maximized over equals
a one-shot variation on a free-energy difference:
The Helmholtz free energy is defined,
in conventional thermodynamics, as $F := E - TS$.
The eigenenergy $\Energy_\mu(0)$
replaces the average energy $E$.
$p_\mu$ equals a probability, so
$- \log ( p_\mu )$ equals a surprisal:
Consider preparing $\ket{\psi_1} \ket{ \varphi {=} 0 }$,
then measuring the energy.
The surprisal quantifies the information you gain,
or the surprise you register, upon learning the outcome.
Averaging the surprisal over many trials
yields the Shannon entropy, 
$S_\Sh = - \sum_{\mu = \pm }  p_\mu  \log  p_\mu$.
The Shannon entropy is proportional to 
the thermodynamic entropy $S$,
for equilibrium states.
Hence the $- \frac{1}{\beta} \log ( 1 / p_\mu )$ 
is a one-shot variation on the $- TS$ in $F$.
The equilibrium state $\exp( - \beta H_\mol ) / Z_\mol$ 
has a free energy of
$F = - \frac{1}{\beta} \log Z_\mol$.
Hence the minimal one-shot work has the form
(one-shot nonequilibrium free energy) - (equilibrium free energy).

Equation~\eqref{eq_W_Min_Gen_Result} reduces 
to the expected formula in a simple case.
Suppose that the diabatic state $\ket{\psi_1} \ket{ \varphi {=} 0 }$ exactly equals 
the energy eigenstate $\ket{ \Energy_+(0) } \ket{ \varphi {=} 0 }$
and that the equilibrium state $e^{ - \beta H_\mol } / Z_\mol$
exactly equals the energy eigenstate 
$\ket{ \Energy_- (0) }  \ket{ \varphi {=} 0 }$.
First, the final-state probabilities become $p_+ = 1$ and $p_- = 0$.
Therefore, the maximum evaluates to $\Energy_+ (0)$.
Second, the Boltzmann weight $e^{ - \beta \Energy_- (0) } / Z_\mol = 1$.
Therefore, the equilibrium free energy reduces to
$- \frac{1}{\beta} \log Z_\mol  =  \Energy_- (0)$.
Substituting into Eq.~\eqref{eq_W_Min_Gen_Result} yields 
a $W_\Min$ of $\Energy_+ (0)  -  \Energy_- (0)$.
The minimal work required to photoexcite 
the ground \emph{cis} state into the excited \emph{cis} state
equals the energy gap.

\end{appendices}

%
%
\bibliographystyle{h-physrev}
\bibliography{Isomer_RT_Refs}

\end{document}